\documentclass{article}

\usepackage{arxiv}

\usepackage[utf8]{inputenc} 
\usepackage[T1]{fontenc}    
\usepackage[hypertexnames=false]{hyperref}       
\usepackage{url}            
\usepackage{booktabs}       
\usepackage{amsfonts}       
\usepackage{nicefrac}       
\usepackage{microtype}      
\usepackage{graphicx}
\usepackage{natbib}
\usepackage{doi}
\usepackage{array}
\usepackage{multirow}
\usepackage{mathtools}
\usepackage{amsmath}
\usepackage{adjustbox}
\usepackage{afterpage}
\usepackage{icomma}
\usepackage{nicematrix}
\usepackage{xltabular}
\usepackage{caption}

\DeclareMathOperator{\E}{\mathbb{E}}

\title{Visualizing Joint Ranking Distributions in Network Meta-Analysis with Hammock Plots}

\author{ \href{https://orcid.org/0009-0009-2270-8943}{\includegraphics[scale=0.06]{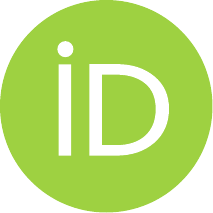}\hspace{1mm}Carmen Alicia Liang Zhen} \\
	Department of Statistics and Actuarial Science\\
	University of Waterloo\\
	Waterloo, Ontario\\
    N2L 3G1\\
    Canada\\
	\And
	\href{https://orcid.org/0000-0003-4124-2498}{\includegraphics[scale=0.06]{orcid.pdf}\hspace{1mm}Audrey B{\'e}liveau} \thanks{Corresponding Author: \texttt{audrey.beliveau@uwaterloo.ca}} \\
	Department of Statistics and Actuarial Science\\
	University of Waterloo\\
	Waterloo, Ontario\\
    N2L 3G1\\
    Canada\\
    \And
    \href{https://orcid.org/0009-0007-3651-9268} {\includegraphics[scale=0.06]{orcid.pdf}\hspace{1mm}Matthias Schonlau} \\
	Department of Statistics and Actuarial Science\\
	University of Waterloo\\
	Waterloo, Ontario\\
    N2L 3G1\\
    Canada\\
}

\date{}

\renewcommand{\headeright}{}
\renewcommand{\undertitle}{}
\renewcommand{\shorttitle}{Visualizing Joint Ranking Distributions in NMA with Hammock Plots}

\hypersetup{
pdftitle={Visualizing Joint Ranking Distributions in Network Meta-Analysis with Hammock Plots},
pdfauthor={Carmen Alicia Liang Zhen, Audrey B{\'e}liveau, Matthias Schonlau},
pdfkeywords={hammock plots; network meta-analysis; parallel coordinate plots; posterior uncertainty; ranking metric; treatment hierarchies},
}

\begin{document}
\maketitle

\begin{abstract}
Treatment ranking  provides a summary of the relative effectiveness of competing treatments and is a key output of network meta-analysis (NMA). Existing visualizations primarily summarize marginal ranking distributions or present individual treatment hierarchies with limited insight into uncertainty in the overall hierarchy. We introduce hammock plots as a visualization framework for the joint posterior distribution of treatment rankings. Although motivated by treatment ranking in NMA, the proposed framework is broadly applicable to any setting in which uncertainty in posterior rankings is of interest. Specifically, we develop separate hammock plots to display relationships between (a) adjacent treatments or (b) adjacent ranks. Using highlighting, we show how to assess posterior support for selected treatment hierarchies, to evaluate hierarchies identified by ranking metrics against the posterior distribution, and to examine uncertainty within selected regions of the treatment hierarchy. We apply the proposed visualizations to a thrombolytic treatment network, which revealed patterns in the posterior ranking distribution and the location of uncertainty in treatment hierarchies that were difficult to identify using existing ranking graphics. We also provide a  web application to generate all of the proposed hammock plot variations for any data set.
\end{abstract}


\section{Introduction}
\label{s:intro}

Treatment ranking is often a central goal of comparative effectiveness research. When multiple interventions are available for a given condition, clinicians, patients, guideline developers, and policy makers are interested in how the available options compare overall. But as the number of competing treatments increases, interpretation becomes increasingly challenging because of the large number of pairwise comparisons. Ranking provides a convenient way to summarize this complexity by translating numerous treatment comparisons into an ordered assessment of treatment performance.

Network Meta-Analysis (NMA) provides a natural framework for treatment ranking because it enables the simultaneous comparison of multiple interventions by combining direct and indirect evidence across a network of studies. Direct evidence comes from within-trial treatment comparisons, while indirect evidence uses a common comparator to evaluate treatments that have never been compared within the same study \citep{lumley2002network}. After fitting an NMA model to the data, relative treatment effects are estimated. To characterize the uncertainty in these estimates, samples of relative effects can be generated through Markov Chain Monte Carlo (MCMC) simulations in a Bayesian framework or parametric bootstrap in a frequentist framework \citep{rucker2015ranking}. For the sake of simplicity, we adopt Bayesian terminology throughout this paper, but the concepts discussed are equally applicable to frequentist NMA settings.
Summary quantities of interest such as an overall treatment ranking and marginal posterior ranking probabilities can be derived from these samples. To communicate these results, a variety of graphical tools have been developed in the context of NMA. Although published NMAs often consider multiple outcomes, these outcomes are typically analyzed separately, and we focus on visualizing treatment rankings for an individual outcome. Methods for summarizing treatment rankings across multiple outcomes, including radar plots \citep{stafoggia2011spie}, scatterplots \citep{chaimani2013graphical}, rank-heat plot \citep{veroniki2016rank}, spie charts \citep{daly2020spie}, and beading plots \citep{chen2024beading}, address a complementary but distinct objective.

A common class of visualizations for individual outcomes focuses on the marginal distribution of treatment ranks. Rankograms display the rank probability distribution of each treatment as a separate line chart or barchart \citep{salanti2011graphical}, while stacked barcharts and pie charts display the same probabilities as segments that sum to one \citep{veroniki2016rank}. Bubble plots display ranking probabilities in a matrix-like layout, using both numerical values and circles whose areas are proportional to the corresponding probabilities \citep{veroniki2016rank}. Cumulative ranking probability plots instead summarize the cumulative ranking distribution by displaying the probability that treatment $i$ is ranked within the top $r$ treatments for each rank $r$ \citep{salanti2011graphical}.

Additional graphical tools for individual outcomes focus on displaying a treatment hierarchy by ordering treatments from the best to the worst using a ranking metric.
Rank charts arrange treatments vertically and use a sequential color palette to convey their relative rank \citep{tan2014novel}, while radial plots position treatments on concentric circles, with a greater distance from the center generally indicating better performance \citep{nevill2023multifaceted}. Another plot used by \citet{chiocchia2023complexity} compares hierarchies identified by multiple ranking metrics within a single plot, where treatments are represented by uniquely colored circles and the same treatments are connected across hierarchies to track changes in ranks.

All of these visualization tools provide useful but partial summaries of treatment ranking. The first group of graphs summarizes the probability of an individual treatment obtaining a given rank, but they do not capture the joint distribution of treatment ranks. From these plots, we can answer questions such as the probability that ``treatment $A$ is ranked first'', or alternatively the probability that ``treatment $B$ is ranked second''. We cannot answer joint questions such as the probability that ``treatment $A$ is ranked first and treatment $B$ is ranked second''. This limitation is important because treatment ranks are dependent, that is improving the rank of one treatment necessarily affects the ranks of the others. Therefore, the probability that ``treatment $A$ is ranked first and treatment $B$ is ranked second'' is not simply the product of their respective marginal probabilities.
Moreover, the second group of graphs summarizes NMA results into a hierarchy based on a ranking metric. However, these graphs do not convey how likely a particular hierarchy is among all possible rank permutations. They also do not communicate how certain the relative ordering of treatments is or how confidently one treatment can be reported to be superior to another. Beyond these graphical limitations,  the choice of an appropriate metric is not always straightforward because each metric is designed to address different ranking questions and therefore may produce different treatment hierarchies \citep{chiocchia2020agreement}. The resulting hierarchy can be influenced by several factors, including imbalances in the number of studies across treatments, network configuration, and differences in effect sizes between interventions \citep{davies2021degree, kibret2014bias}. Thus, a hierarchy based on a single ranking metric may oversimplify the posterior ranking distribution, and it should be interpreted with caution. 

To address these gaps, this paper proposes the use of parallel coordinates --- specifically, hammock plots \citep{schonlau2003visualizing, schonlau2024hammock} ---
to visualize the joint posterior distribution of treatment rankings. 
Parallel coordinate plots \citep{inselberg1985plane,wegman1990} use parallel axes (rather than perpendicular axes like the scatter plot) to display data. 
Hammock plots were the first plots with parallel axes to also display relationships among categorical variables. 
To our knowledge, this is the first application of hammock plots for visualizing joint ranking distributions.
Here, we extend the use of hammock plots to the NMA setting, where observation counts are replaced by posterior proportions, and variables represented on the axes are either treatments or ranks. By connecting categories across adjacent axes, hammock plots simultaneously display marginal ranking probabilities, bivariate rank relationships, and optionally uncertainty in complete ranking hierarchies. Although we focus on NMA, the proposed methods are applicable in any context when samples are available, either from the joint posterior ranking distribution of treatments (Bayesian) or from the joint distribution of estimated ranks (frequentist).

Section \ref{s:background} introduces notation used in Section \ref{s:Method}, which develops the five proposed hammock plot methods. Section \ref{s:Application} illustrates these methods through an NMA of eight thrombolytic treatments, revealing the new insights they offer. The paper concludes with a discussion in Section \ref{s:Discussion}.

\section{Notation}\label{s:background}
Let $T$ denote the total number of treatments in the network, and let $S_T$ denote the set of all treatments. Bayesian NMA produces $N$ posterior samples of the relative effects for $T-1$ treatments relative to a reference treatment, with the relative effect of the reference treatment fixed at zero (e.g., on the log odds ratio or mean difference scale). Within each posterior sample, treatments in $S_T$ can be ordered according to their relative treatment effects. Every treatment is assigned a unique rank and every rank is occupied by exactly one treatment in a sample. We define the function $Rank(\cdot)$ which maps a treatment to its corresponding rank. For a treatment $t$, $Rank(t)$ denotes the rank this treatment takes, with a range of 1 to $T$, where rank 1 is the best and rank $T$ is the worst. Consequently, $p_{tr} = P\{Rank(t) = r|data\}$ denotes the posterior probability that treatment $t$ is assigned the rank $r$. This probability is estimated by the relative frequency that treatment $t$ occupies the rank $r$ across the $N$ posterior samples:
$p_{tr} = \frac{1} {N} \sum_{n=1}^N \mathbb{I}\{Rank_n(t)=r\},$
where $\mathbb{I}(A)$ is the indicator function which equals $1$ if the statement $A$ is true and $0$ otherwise, and $Rank_n(t)$ denotes the rank of treatment $t$ in sample $n$.

Similarly, we define $Treatment(\cdot)$ which maps a rank to the treatment assigned to this rank. Then $Treatment(r)$ and $P\{Treatment(r) = t|data\}$ output the treatment name assigned to rank $r$ and the posterior probability that rank $r$ is occupied by treatment $t$ respectively. This posterior probability is estimated by
$P\{Treatment(r) = t|data\} = \frac{1} {N} \sum_{n=1}^N \mathbb{I}\{Treatment_n(r)=t\},$
where $Treatment_n(r)$ is the treatment occupying rank $r$ in posterior sample $n$. Since each posterior sample defines a one-to-one correspondence between treatments and ranks, we have
$p_{tr}=P\{Rank(t) = r|data\} = P\{Treatment(r) = t|data\}.$

Ordering the treatments in each posterior sample produces a complete treatment ranking and a complete treatment hierarchy. We define a complete treatment ranking for a fixed treatment ordering $(t_1,...,t_T)$ as
$R=[Rank(t_1), ..., Rank(t_T)],$
where the $i$th element is the rank that treatment $t_i$ takes. Likewise, a complete treatment hierarchy is
$H=[Treatment(1),$ $..., Treatment(T)]$
where the $r$th element denotes the treatment assigned to rank $r$.

\section{Hammock Plot Methods}\label{s:Method}

In this section, we develop a visualization framework for exploring the posterior distribution of treatment rankings in NMA using hammock plots. We begin by showing how a stacked barchart can be extended to a hammock plot in Section \ref{subsec:Bivariate_distribution} to visualize pairwise relationships between adjacent treatments or adjacent ranks. Section \ref{subsec:Highlighting} describes how highlighting can be used to display the joint distribution of posterior ranks and the uncertainty in selected treatment hierarchies. Finally, to improve interpretability in large treatment networks, we propose methods for visualizing specific subsets of treatments in Section \ref{subsec:subsetting}. Web Table~\ref{tab:hammock_plots} provides an overview of the five hammock plot methods introduced in this section.

Throughout Section \ref{s:Method}, we use a previously published NMA dataset comprising six diabetes treatments as a running example \citep{elliott2007incident, dias2011nice}. The dataset is analyzed using a Bayesian random-effects model with a binomial likelihood and a complementary log-log link, generating $N=19,998$ posterior samples of relative treatment effects from three chains after thinning \citep{beliveau2019bugsnet, dias2011nice}. The first seven posterior samples of relative treatment effects for the diabetes dataset are provided in Web Table~\ref{tab:relative_effects}.

Within each posterior sample, treatments are assigned ranks according to their relative treatment effects to obtain a complete treatment ranking $R$ and the corresponding hierarchy $H$. All complete treatment rankings are stored in the rank matrix (see Web Table~\ref{tab:rank_matrix}), and the corresponding  treatment hierarchies are stored in the treatment matrix (see Web Table~\ref{tab:treatment_matrix}). In both matrices, each row corresponds to one posterior sample. These two matrices form the basis for constructing the hammock plots described in the following subsections.

\subsection{Visualizing Bivariate Posterior Treatment Rankings}\label{subsec:Bivariate_distribution}

A hammock plot extends the stacked barchart by connecting adjacent ranks or treatments \citep{schonlau2003visualizing, schonlau2024hammock}. In NMA, stacked barcharts are widely used to summarize the posterior probability that each treatment occupies a specific rank \citep{veroniki2016rank}. A description and illustration of the stacked barcharts for the diabetes dataset is provided in \hyperlink{WebAppendixA}{Web Appendix A}.

\begin{figure*}[htb]
    \centerline{\includegraphics[width=\linewidth, trim=0 15 15 0, clip]{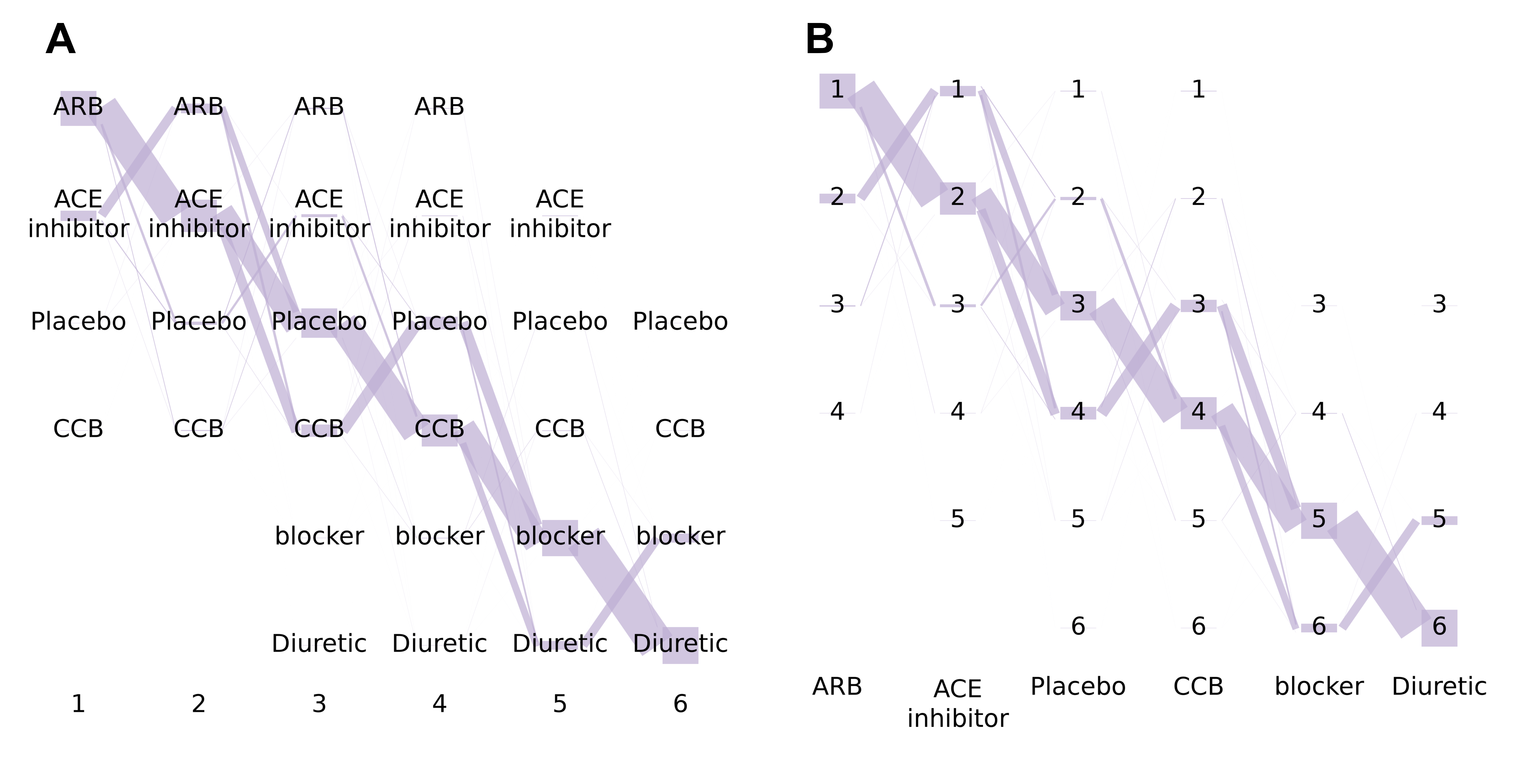}}
    \caption{Two hammock plots of the posterior distribution for the diabetes network. (A) Each column presents the posterior probability distribution of treatments at a fixed rank, constructed from the treatment matrix. (B) Each column represents the posterior probability distribution of ranks for a fixed treatment, constructed from the rank matrix. Treatments are ordered by SUCRA, from top to bottom within each rank axis of (A) and from left to right across the column labels of (B). In both charts, the height of each segment is proportional to the corresponding posterior probability. The width of the connector is proportional to the posterior probability that two treatments are assigned to two ranks at the same time.}
    \label{fig:Figure2_simpleHammock}
\end{figure*}

Figure~\ref{fig:Figure2_simpleHammock} displays two hammock plots for the diabetes dataset: they consist of a sequence of parallel axes (one in each column), each representing a rank (Figure~\ref{fig:Figure2_simpleHammock}A) or a treatment (Figure~\ref{fig:Figure2_simpleHammock}B). 
On each parallel axis, the height of each stacked segment is proportional to the corresponding posterior probability. Therefore, the hammock plots preserve the information displayed in the stacked barcharts in Web Figure~\ref{fig:Figure1_snapshot}.
Unlike in a stacked barchart, the columns of segments on each axis of a hammock plot are not closely stacked; they are separated by gaps. The gaps are needed to align  the category labels across columns; they do not change the marginal posterior distributions on each axis.

Additionally, a missing treatment segment on a rank axis or a missing rank segment on a treatment axis indicates zero posterior probability (i.e. $p_{tr} = 0$). Therefore, each axis demonstrates either the marginal distribution of treatments at a fixed rank or the marginal distribution of ranks for a fixed treatment.

In a hammock plot, each stacked segment is called a univariate bar, and the rectangular boxes that links categories on the adjacent axis are called connectors. In Figure~\ref{fig:Figure2_simpleHammock}A, each connector joins a pair of treatments $t_1$ and $t_2$ on adjacent rank axes $r$ and $r+1$. The width of the connector is proportional to the posterior probability that $t_1$ is assigned rank $r$ and $t_2$ is assigned rank $r+1$ in the same posterior sample, that is: $P\{Treatment(r)=t_1, Treatment(r+1)=t_2|data\}=\frac{1}{N}\sum_{n=1}^{N}
\mathbb{I}\!\left\{Treatment_n(r)=t_1, Treatment_n(r+1)=t_2 \right\}$.
A wider connector indicates that the corresponding pair of treatments occupies the corresponding ranks more frequently in the posterior samples. Here, the width of a connector refers to the distance between two longer parallel sides, rather than the vertical distance. Additionally, the widths of all connectors leaving one univariate bar partition the height of that bar. Since the height of the univariate bar for $Treatment(r)=t_1$ is proportional to $P\{Treatment(r)=t_1|data\}$, then it satisfies
$\sum_{t_2 \in S_T \backslash \{t_1\} } P\{Treatment(r)=t_1, \ Treatment(r+1)=t_2|data\} = P\{Treatment(r)=t_1|data\}.$ Therefore, all connectors between two axes visualize the complete bivariate posterior treatment distribution among two ranks. Web Figure~\ref{fig:Figure12_Hammock_annotation} provides an annotated summary of these graphical elements and their interpretations.

Similar interpretations apply to Figure~\ref{fig:Figure2_simpleHammock}B. Each connector joins ranks on adjacent treatment axes. The width of the connector is proportional to the posterior probability that treatment $t_1$ takes rank $r_1$ and treatment $t_2$ occupies rank $r_2$ in the same posterior sample, namely $P\{Rank(t_1)=r_1, \ Rank(t_2)=r_2|data\}=\frac{1}{N}\sum_{n=1}^{N}
\mathbb{I}\!\left\{
Rank_n(t_1)=r_1, \ 
Rank_n(t_2)=r_2
\right\},$
which is equivalent to
$P\{Treatment(r_1)=t_1, Treatment(r_2) =t_2|data\}.$
Additionally, the connectors connecting the univarate bar of $Rank(t_1)=r_1$ partition its margin posterior probability:
$\sum_{r_2=1, r_2\neq r_1}^T P\{Treatment(r_1)=t_1, Treatment(r_2) =t_2|data\} = P\{Treatment(r_1)=t_1|data\}.$
Thus, the connectors between two treatment axes represent the complete joint posterior distribution of the ranks occupied by the two treatments.

Figure~\ref{fig:Figure2_simpleHammock}A is constructed using the treatment matrix with ranks as axes, and Figure~\ref{fig:Figure2_simpleHammock}B is constructed using the rank matrix with treatments as axes. The one-to-one correspondence between the treatments and ranks in these two input datasets allows the univariate bars in the two hammock plots to display the same marginal posterior probabilities. However, it is important to note that the two plots emphasize different bivariate relationships. Figure~\ref{fig:Figure2_simpleHammock}A fixes adjacent ranks and visualizes the joint posterior distribution of the treatments occupying those ranks, while Figure~\ref{fig:Figure2_simpleHammock}B fixes adjacent treatments and visualizes the joint posterior distribution of ranks occupied by those treatments.

At first glance, the two hammock plots may appear to be transpositions of each other, similar to the two stacked barcharts in Web Figure~\ref{fig:Figure1_snapshot}, where the horizontal arrangement of stacked segments in one chart becomes the vertical arrangement in the other. However, because connectors are drawn only between adjacent axes, a bivariate relationship displayed directly in one plot may not be displayed directly in the other plot. For example, the joint event $Treatment(1)=\text{ARB} \ \text{and}\ Treatment(2) = \text{Placebo}$ is explicitly shown by a single connector in Figure~\ref{fig:Figure2_simpleHammock}A. Its equivalent event, $Rank(\text{ARB})=1 \ \text{and} \ Rank(\text{Placebo})=2$ is not displayed in Figure~\ref{fig:Figure2_simpleHammock}B. This is because the ARB and Placebo axes are not adjacent. Although reordering the treatment axes would make this connector visible, it would simultaneously separate another pair of treatments and remove the corresponding connectors.

\subsection{Highlighting Treatment Hierarchies}\label{subsec:Highlighting}

The previous section demonstrated how a hammock plot visualized bivariate posterior relationships between two adjacent ranks or treatments. We now extend our focus to the joint posterior distributions of all ranks and all treatments on selected paths. Each posterior sample is a realization from the joint distribution and corresponds to a complete path in a hammock plot. Each path represents either a complete treatment hierarchy or a complete treatment ranking. Individual paths are not easily identified and compared because each univariate bar is connected to multiple connectors, and many paths share the same connectors.
To address this limitation, we extend the hammock plot by highlighting complete treatment hierarchies and rankings.

\subsubsection{Highlighting Frequent Paths}
\begin{figure*}[htb]
    \centerline{\includegraphics[width=1\textwidth, trim=10 10 15 0, clip]{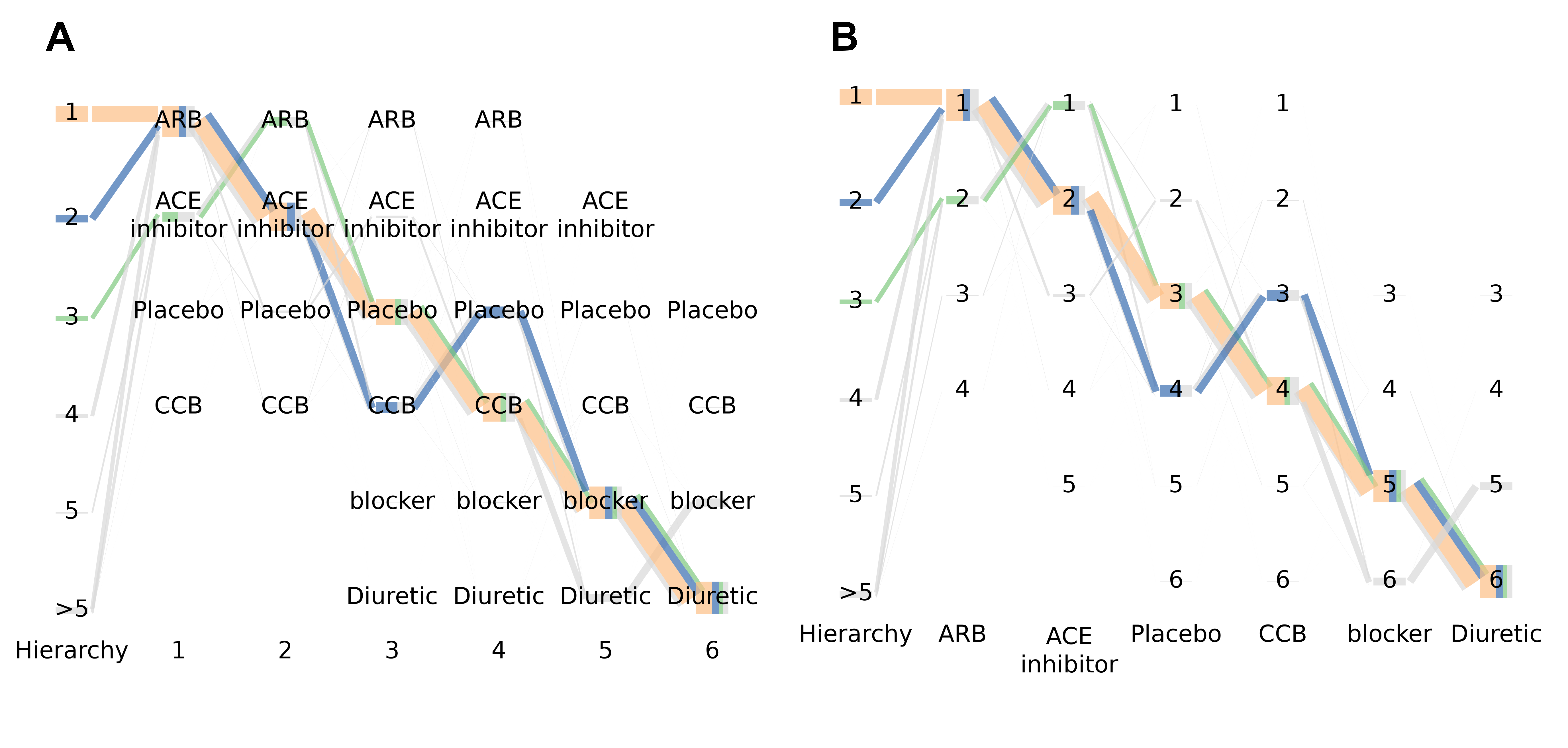}}
    \caption{Two Hammock plots highlighting frequency for the diabetes dataset. (A) Each column presents the posterior probability distribution of treatments at a fixed rank, constructed from the treatment matrix. (B) Each column represents the posterior probability distribution of ranks for a fixed treatment, constructed from the rank matrix. Treatments are ordered by SUCRA, from top to bottom within each rank axis of (A) and from left to right across the column labels of (B). In both plots, the additional categorical variable ``Hierarchy" classifies paths into different levels based on the frequencies. Hierarchy level 1 corresponds to the most probable path, while higher levels represents less probable paths. The width of each path is proportional to the posterior probability of the complete path.}
    \label{fig:Figure3_Frequency}
\end{figure*}

We first highlight the most frequently observed paths in the posterior distribution. Figure~\ref{fig:Figure3_Frequency} illustrates the resulting highlighted hammock plots for the diabetes dataset. 
Figure~\ref{fig:Figure3_Frequency}A uses ranks as parallel axes and Figure~\ref{fig:Figure3_Frequency}B uses treatments. These plots extend the hammock plots in Figure~\ref{fig:Figure2_simpleHammock} by introducing an additional parallel axis in the front labeled ``Hierarchy''. ``Hierarchy" is a categorical variable. It has levels 1 to 5 and an additional category `$>5$'. Hierarchy 1 corresponds to the most probable path in the posterior samples, Hierarchy 2 is the second most frequent path, and so on, while `$>5$' groups all paths outside of them. For visual clarity, only the three most probable paths are highlighted using distinct colors, while the remaining paths are shown in gray.

To construct Figure~\ref{fig:Figure3_Frequency}A, we calculate the frequency of each unique treatment hierarchy in the treatment matrix. The corresponding hierarchy levels are then added to each sample in the treatment matrix as the categorical variable ``Hierarchy". Recall that $H=[Treatment(1), ..., Treatment(T)]$ denote a complete treatment hierarchy. Then each posterior sample in the treatment matrix is one realization from the multivariate posterior distribution of $H$. The posterior probability of a hierarchy $h$ is estimated by the proportion of posterior samples that satisfies $H=h$, and is given by
$P\{H=h|data\} =\frac{1}{N} \sum_{n=1}^N \mathbb{I}(h_n=h)$,
where $h_n$ is the treatment hierarchy in posterior sample $n$. In the highlighted hammock plot, this posterior probability is represented  by the width of the corresponding highlighted path. Specifically, the connectors and univariate bars highlighted with the same color belong to the same path and have identical width which is proportional to $P\{H=h|data\}$. Therefore, the width of each highlighted path remains constant from the first axis to the last axis, and wider highlighted paths correspond to more probable treatment hierarchies. Comparing the relative width of one highlighted path with the others provides a visual summary of the posterior uncertainty associated with the complete treatment hierarchy. Particularly, a dominant path that is substantially wider than the others indicates lower uncertainty.

Figure~\ref{fig:Figure3_Frequency}B is constructed in a similar way to Figure~\ref{fig:Figure3_Frequency}A, but using the rank matrix. The graphical interpretations for Figure~\ref{fig:Figure3_Frequency}A apply equivalently to Figure~\ref{fig:Figure3_Frequency}B, except that each highlighted path now represents a complete ranking $R$. Each posterior sample in the rank matrix is then a realization from the posterior distribution of $R$. The width of the highlighted path is proportional to
$P\{R=r|data\} = \frac{1}{N} \sum_{n=1}^N \mathbb{I} (r_n=r),$
where $r$ is the observed complete ranking and $r_n$ denotes the treatment ranking in posterior sample $n$. Since there is a one-to-one correspondence between the treatment matrix and the rank matrix, every complete hierarchy $h$ uniquely identifies a complete ranking $r$ and vice versa. Therefore,
$P(H=h|data)=P(R=r|data),$
where $r$ is the ranking corresponding to the hierarchy $h$. The path of one particular ``Hierarchy" level in Figure~\ref{fig:Figure3_Frequency}A has width identical to the path in the same ``Hierarchy" level of Figure~\ref{fig:Figure3_Frequency}B, even though the two plots have different graphical structures.

\subsubsection{Highlighting Paths from Different Ranking Metrics}

The frequency-based highlighting strategy emphasizes the most frequently observed paths in the posterior distribution. An alternative approach is to highlight the treatment hierarchies identified by ranking metrics. The ranking metrics that we will consider are surface under the cumulative ranking curve (SUCRA), the probability of being the best (PBV), the expected treatment effect (EV), and the posterior median rank. The SUCRA of a treatment $t$ is calculated by $[T-\E\{Rank(t)|data\}]/(T-1)$, where $\E\{Rank(t)|data\}$ denotes the posterior mean rank of $t$ \citep{nevill2023multifaceted}. Higher SUCRA values indicate better treatments. Since SUCRA is a linear transformation of the posterior mean rank with a negative slope, treatments with lower posterior mean rank always have higher SUCRA values. Therefore, the two metrics always suggest the same treatment hierarchy, and we use SUCRA to represent both metrics. PBV is calculated by $P\{Rank(t)=1|data\}$ for all possible treatment $t$. Treatments with larger PBV are ranked higher in the treatment hierarchy. The EV orders treatments according to their posterior mean relative treatment effects estimated from MCMC samples. Depending on the outcome of interest, either smaller or larger posterior mean relative treatment effects indicates better treatments. Similarly, the posterior median rank orders treatments according to the median of the posterior rank distribution.

\begin{figure*}[htb]
    \centerline{\includegraphics[width=1\textwidth]{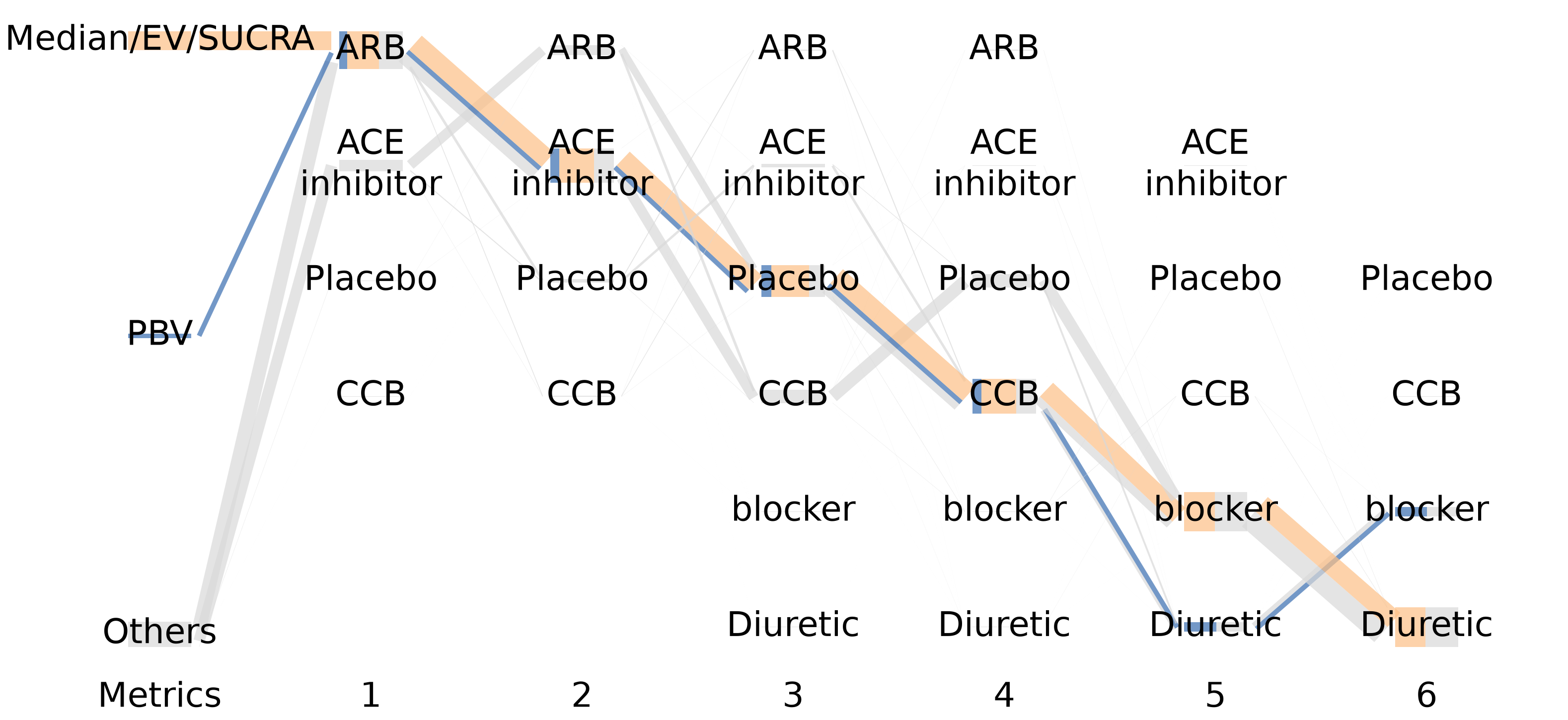}}
    \caption{Hammock plot highlighting ranking metrics for the diabetes dataset. Ranks are used as the parallel axes and an additional variable ``Metrics" groups posterior samples according to treatment hierarchies identified by selected ranking metrics. The width of each highlighted path is proportional to the posterior probability of the corresponding treatment hierarchy. Treatments are ordered by SUCRA, from top to bottom within each rank axis. Categories on the Metrics axis are ordered by the number of posterior samples in each category, from largest to smallest, with ``Others" always positioned last.}
    \label{fig:Figure4_metrics}
\end{figure*}

Figure~\ref{fig:Figure4_metrics} is an extension of Figure~\ref{fig:Figure2_simpleHammock}A with an additional parallel axis ``Metrics". This plot
highlights the hierarchies identified by the selected ranking metrics. Similar to Figure~\ref{fig:Figure3_Frequency}A, this plot is constructed using the treatment matrix, but the ``Metrics'' axis has categories based on ranking metrics rather than frequency levels. Ranking metrics that produce the same treatment hierarchy are grouped into a single category, with their names separated by a slash (/). This avoids highlighting the same hierarchy multiple times. For example, the category ``Median/EV/SUCRA" in Figure~\ref{fig:Figure4_metrics} indicates that these three metrics generate the same treatment hierarchy. Posterior samples in the treatment matrix are then assigned to the corresponding metric category if their treatment hierarchies match the metric-based hierarchy; otherwise they are classified into the category ``Others".

The interpretation of Figure~\ref{fig:Figure4_metrics} is analogous to that of Figure~\ref{fig:Figure3_Frequency}A. Suppose $m$ is a ranking metric, and let $h_m=[Treatment(1), ..., Treatment(T)]$ denote the treatment hierarchy generated by metric $m$. The width of the highlighted path corresponding to the metric $m$ is proportional to $P\{H=h_m|data\}$, with a wider path indicating greater posterior support for that hierarchy. Since the hammock plot displays all possible treatment hierarchies sampled from the posterior joint distribution, highlighting allows the hierarchy identified by metric $m$ to be interpreted in the context of the entire distribution. A narrow highlighted path indicates greater uncertainty in the complete treatment hierarchy, as a larger proportion of the posterior distribution is allocated to other hierarchies.

\subsection{Visualizing a Subset of Treatments}\label{subsec:subsetting}
When a network consists of a large number of treatments, hammock plots can become visually cluttered: an increasing number of connectors, paths and univariate bars must be displayed. It becomes difficult to identify the most relevant path and interpret the overall posterior distribution. Additionally, posterior probability may be distributed across many hierarchies, resulting in narrow paths and making highlighted hierarchies less distinguishable from other competing hierarchies.

We propose two approaches in~\hyperlink{WebAppendixB}{Web Appendix B} to reduce visual clutter by subsetting the treatments displayed in a hammock plot. The first approach focuses on treatments that ranked within the top $k$ positions in a hierarchy identified by a specific ranking metric, where $k < T$. Treatments or ranks outside this region are grouped into an ``Others" category. The second approach focuses on a selected subsets of treatments. Treatments outside the selected subsets are completely ignored, including their position in the path. Only the relative ordering of the selected treatments is visualized. Both approaches combine hierarchies that share the same reduced ordering into a single path with width representing their aggregated posterior probability. By doing so, they concentrate the posterior information that is originally distributed across many complete hierarchies and make the posterior structure of interest easier to distinguish.

\section{Application to an NMA of Thrombolytic Treatments}\label{s:Application}

\begin{figure*}[htbp]
    \centerline{\includegraphics[width=\linewidth, trim=0 0 0 0, clip]{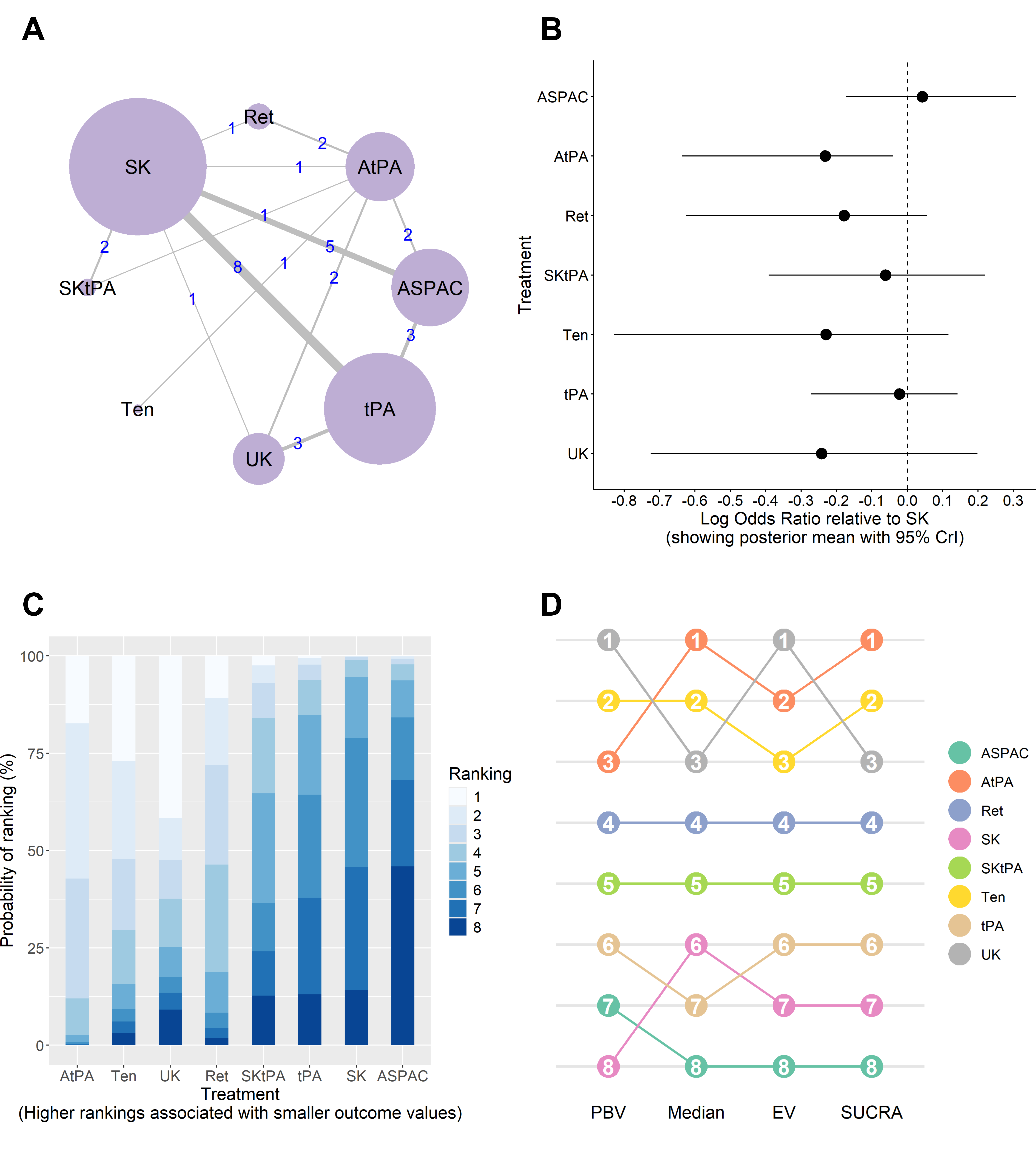}}
    \caption{Summary of the thrombolytic dataset and NMA results. (A) Network plot of the 8 treatments in 28 trials. Each node represents one treatment, where the node size is proportional to the number of studies that examined the corresponding treatment. Edges indicate direct comparisons between connected treatments. The size of the edge is proportional to the number of studies directly comparing the two treatments, with the corresponding number labeled on the edge. (B) Forest plot of the relative treatment effects. Each dot represents the posterior mean log odds ratio for a treatment relative to SK, and horizontal line represents the corresponding 95\% credible interval. (C) Stacked barchart that shows the posterior treatment probabilities at each rank. (D) Treatment hierarchies identified by four commonly used ranking metrics. Circles of the same color represent one treatment. The number within each circle indicates the rank of the corresponding treatment in the hierarchy identified by the ranking metric shown at the bottom.}
    \label{fig:Figure7_thrombo}
\end{figure*}

In this section, we apply the proposed hammock plot methods to reveal important insights into the treatment hierarchy that go beyond those afforded by existing visualization methods. We re-examine the thrombolytic treatment network, which has served as a benchmark example in the development and application of NMA methods following the publication of the original systematic review by \citet{boland2003early}. Specifically, we analyze the published aggregate-data NMA from the R package \texttt{gemtc} \citep{gemtc}, which includes 28 studies comparing eight thrombolytic treatments following acute myocardial infarction: anistreptilase (ASPAC), accelerated alteplase (AtPA), reteplase (Ret), streptokinase (SK), streptokinase plus alteplase (SKtPA), tenecteplase (Ten), alteplase (tPA), urokinase (UK) \citep{lu2006assessing}. The outcome is mortality after 30-35 days, and Figure~\ref{fig:Figure7_thrombo}A presents the network plot for this outcome.

The data is analyzed using a Bayesian random-effects contrast-likelihood NMA model with binomial likelihood function, a logit link, and SK as the reference treatment. Following the default implementation in the R package \texttt{BUGSnet}, independent normal priors with mean 0 and standard deviation 15$u$ are assigned to the baseline and relative treatment effects, where $u=1.3573$ is the largest maximum likelihood estimate of the treatment effect in individual trials \citep{beliveau2019bugsnet}. A uniform prior on $(0, u)$ is used for the between-study heterogeneity standard deviation. Posterior samples are obtained from 3 MCMC chains, each with $1,000$ adaptions, $1,000$ burn-in iterations, and $20,000$ sampling iterations. A thinning factor of 3 is applied to keep every 3 draws in each chain. This produces $19,998$ samples from the joint distribution of relative treatment effects. Within each sample, treatments are ordered from the lowest to the highest relative treatment effect to obtain ranks 1 through 8. These rankings are used to construct the rank matrix and the corresponding treatment matrix as described in Section~\ref{s:Method}.

Figure~\ref{fig:Figure7_thrombo}B presents a forest plot of the posterior mean log odds ratios for each treatment relative to the reference treatment SK and their corresponding uncertainty. Only AtPA shows evidence of a difference from the reference treatment, SK, as its 95\% credible interval excludes 0. A stacked barchart showing the probability that each treatment occupies each possible rank is presented in Figure~\ref{fig:Figure7_thrombo}C. Treatment UK is most likely to be ranked first because it has the largest rank 1 segment. The treatment hierarchies identified by four commonly used ranking metrics are also displayed in Figure~\ref{fig:Figure7_thrombo}D using a visualization analogous to that utilized by \cite{chiocchia2023complexity}. The ranks assigned to each treatment across the four hierarchies are shown in circles, and each treatment is represented by a unique color and  connected across the plot to track changes in its rank.

\subsection{Hammock Plot Reveals Treatment Clusters}
\label{sec: treatment_clusters}

\begin{figure*}[htb]
    \centerline{\includegraphics[width=\linewidth]{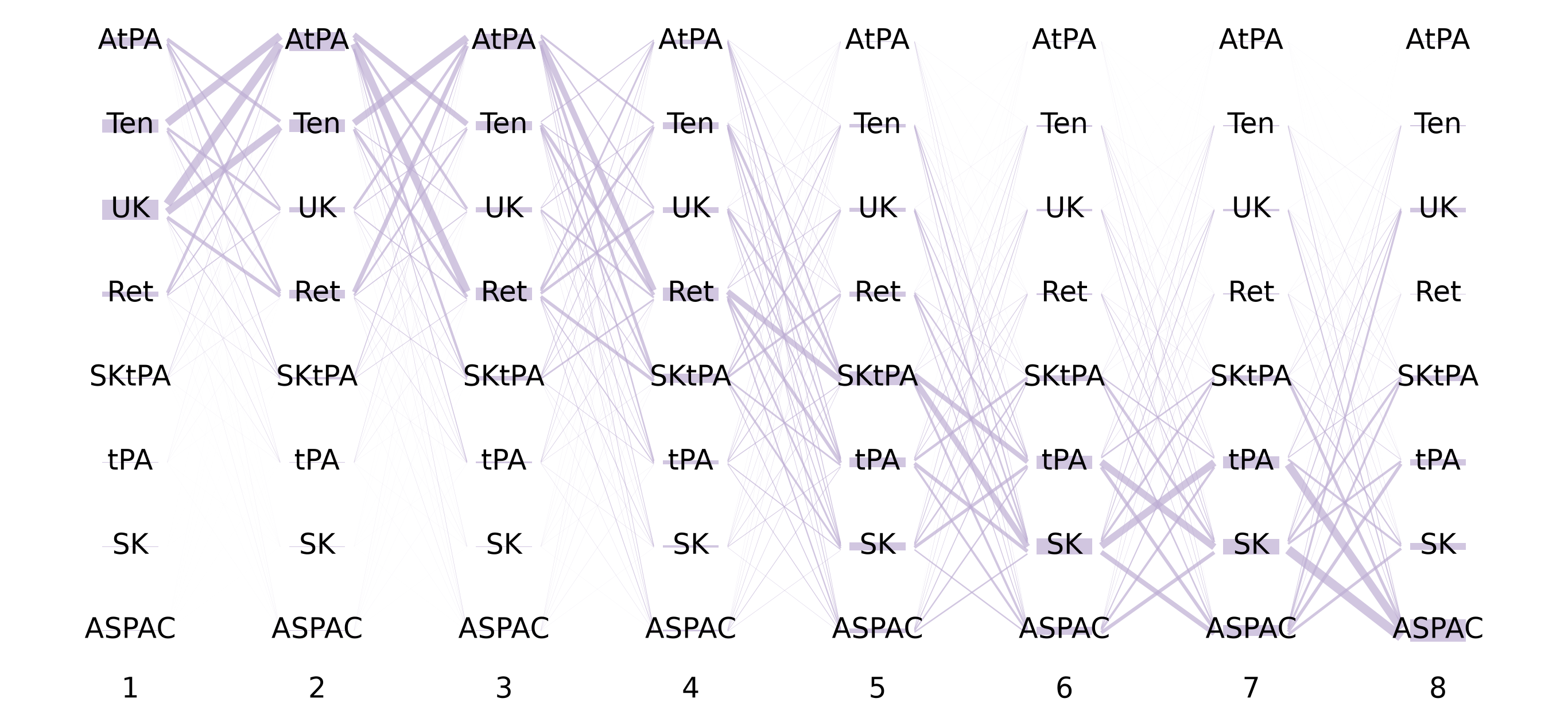}}
    \caption{Standard hammock plot for the thrombolytic dataset. Treatments on each rank axis are ordered by SUCRA. Two broad clusters of high-ranked and low-ranked treatments are apparent, with substantial crossing of connectors within the upper-left and lower-right regions of the plot.}
    \label{fig:Figure8thrombo_simpleHammock}
\end{figure*}

The substantial overlap among the 95\% credible intervals in Figure~\ref{fig:Figure7_thrombo}B suggests that several treatments may have similar effectiveness. However, it is not clear from the forest plot whether this similarity in treatment effects translates to the posterior rankings, or whether any grouping structure emerges in the treatment hierarchy. In addition, the stacked barchart in Figure~\ref{fig:Figure7_thrombo}C shows the probability that each treatment occupies each possible rank, and provides some indication of a separation between treatments that tend to occupy higher or lower ranks. For example, AtPA, Ten, UK and Ret have a high probability of occupying upper ranks. Their bars are dominated by light-colored segments that correspond to ranks 1 to 4, while the remaining treatments are dominated by dark-colored segments that correspond to ranks 5 to 8. However, the stacked barchart is based on marginal rank distributions, and does not provide information about the joint ranking behavior of multiple treatments. The standard hammock plot in Figure~\ref{fig:Figure8thrombo_simpleHammock} supplements these visualizations by displaying all posterior samples simultaneously. This preserves the joint ranking information among treatments, and captures patterns in how treatments are ranked together.

Figure~\ref{fig:Figure8thrombo_simpleHammock} suggests two broad clusters of treatments. The first cluster consists of AtPA, Ten, UK and Ret, which more frequently occupy the top four ranks. This is reflected by the concentration of crossing connectors among the first four ranks in the upper-left region of the plot. The second cluster consists of SKtPA, tPA, SK and ASPAC which more frequently takes the bottom four rank position, as visualized by the numerous crossing connectors in the lower-right corner. Within each cluster, the dense crossing connectors indicate treatments frequently exchange adjacent rank positions. In contrast, the lower-left and upper-right regions contain relatively few connectors, which suggests that treatments rarely move to the opposite end of the treatment hierarchy. A small number of connectors cross the boundary between the two clusters at ranks 4 and 5, indicates the separation is not absolute. Consistent with these observations, the posterior probability that AtPA, Ten, UK and Ret jointly occupy the top four ranks, while SKtPA, tPA, SK and APSAC occupy the bottom four ranks is 45.3\%. Overall, the plot indicates that the posterior uncertainty is concentrated within these two broad groups, rather than spreading uniformly across the entire treatment hierarchy. Although the exact ranks of treatments remain uncertain, treatments tend to remain within the same broad group more often than they exchange between groups. The numerous connectors crossing within each cluster indicate substantial uncertainty in their relative ordering, so no single ordering of treatments within either cluster dominates the posterior distribution. This raises the question of whether the posterior supports any complete treatment hierarchy.

\subsection{Hammock Plot Reveals Absence of Highly Probable Treatment Hierarchies}

\begin{figure*}[htb]
    \centerline{ \includegraphics[width=\linewidth]{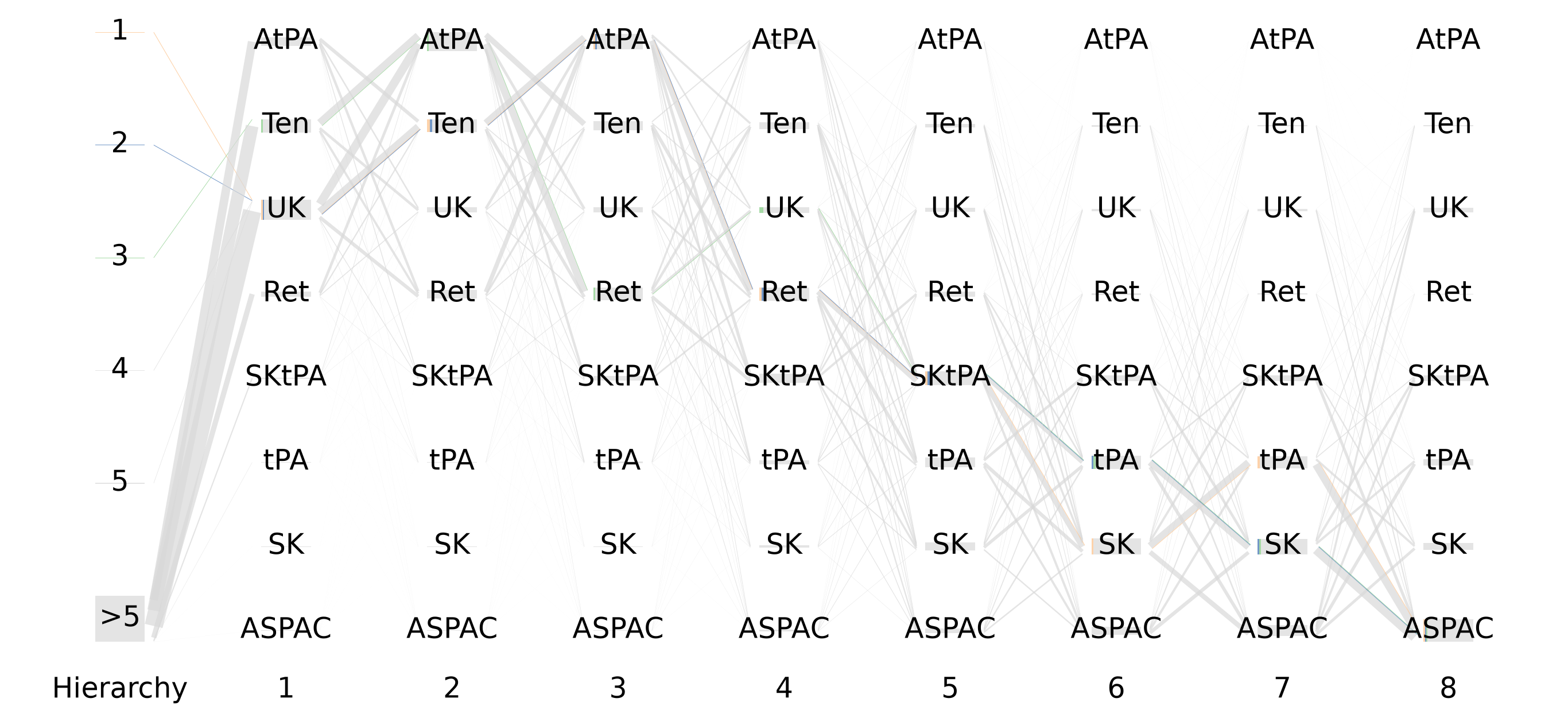}}
    \caption{A hammock plot highlighting frequency for the thrombolytic dataset. Treatments on each rank axis are ordered by SUCRA. The thin highlighted paths indicates that even the most probable hierarchies are not strongly supported by the posterior distribution. Although connector widths could be manually increased to improve the visibility of the highlighted paths, they are intentionally not adjusted so that their widths remain proportional to the posterior probabilities they represent.}
    \label{fig:Figure9_thrombo_fre}
\end{figure*}

Figure~\ref{fig:Figure9_thrombo_fre} presents a hammock plot highlighting frequency for the thrombolytic dataset, but in contrast to the diabetes dataset, no single treatment hierarchy, including the most probable one, dominates the posterior distribution.
We can observe this from the thin highlighted paths in the plot which represent the three most probable hierarchies. In this case, none of the highlighted paths are substantially wider than the others, which equivalently means that none of these hierarchies has a substantially greater relative frequency than the others. Consequently, even the treatment hierarchy with the highest posterior probability is not strongly supported by the posterior distribution. Posterior support is distributed across many competing treatment hierarchies, as reflected by numerous possible paths and crossing connectors in the plot. This indicates high uncertainty in the treatment hierarchy, and many alternative orderings are as plausible as the others. In fact, the most probable hierarchy has a posterior probability of only 1.28\%, followed by 1.18\%, 1.08\%, and 1.05\% for the second, third and fourth most probable hierarchies. All remaining unique treatment hierarchies have posterior probabilities less than 1\%. 

An earlier study by \citet{phillippo2018sensitivity} on the thrombolytic dataset reaches a similar conclusion regarding the uncertainty in treatment recommendations, which highly depend on treatment rankings. Their analysis considered a reduced six-treatment network, excluding UK and APSAC, and focuses on the robustness of the optimal treatment rather than the full treatment hierarchy. Using study-level threshold analysis, they identified several studies were sensitive, and only a very small bias adjustment was required to change the recommended optimal treatment. Similarly, contrast-level threshold analysis found very small thresholds for the comparisons between Ret and AtPA and between Ten and AtPA. Based on this, they conclude that any of these three treatments could reasonably be recommended rather than a single unique optimal treatment.

Figure~\ref{fig:Figure9_thrombo_fre} provides complementary evidence for the conclusion from \cite{phillippo2018sensitivity}. The posterior distribution of treatment hierarchies illustrated in the plot also suggests that the data does not provide clear support for a single treatment recommendation. Although UK is ranked first in three of the four most probable treatment hierarchies, another relatively highly probable hierarchy identifies Ten as the best treatment. Additionally, many of the hierarchies within the ``$>5$" group identify AtPA as the best.
This visualization helps to explain why the optimal treatment recommendation can be sensitive to small changes in evidence: because posterior support is already distributed among multiple plausible hierarchies and none of them receives strong support, relatively small adjustments to the data may be sufficient to change the most probable hierarchy and therefore the preferred treatment.

\subsection{Hammock Plot Reveals Uncertainty Structure of the Metric-Based Treatment Hierarchies}

\begin{figure*}[htb]
\centerline{\includegraphics[width=\linewidth]{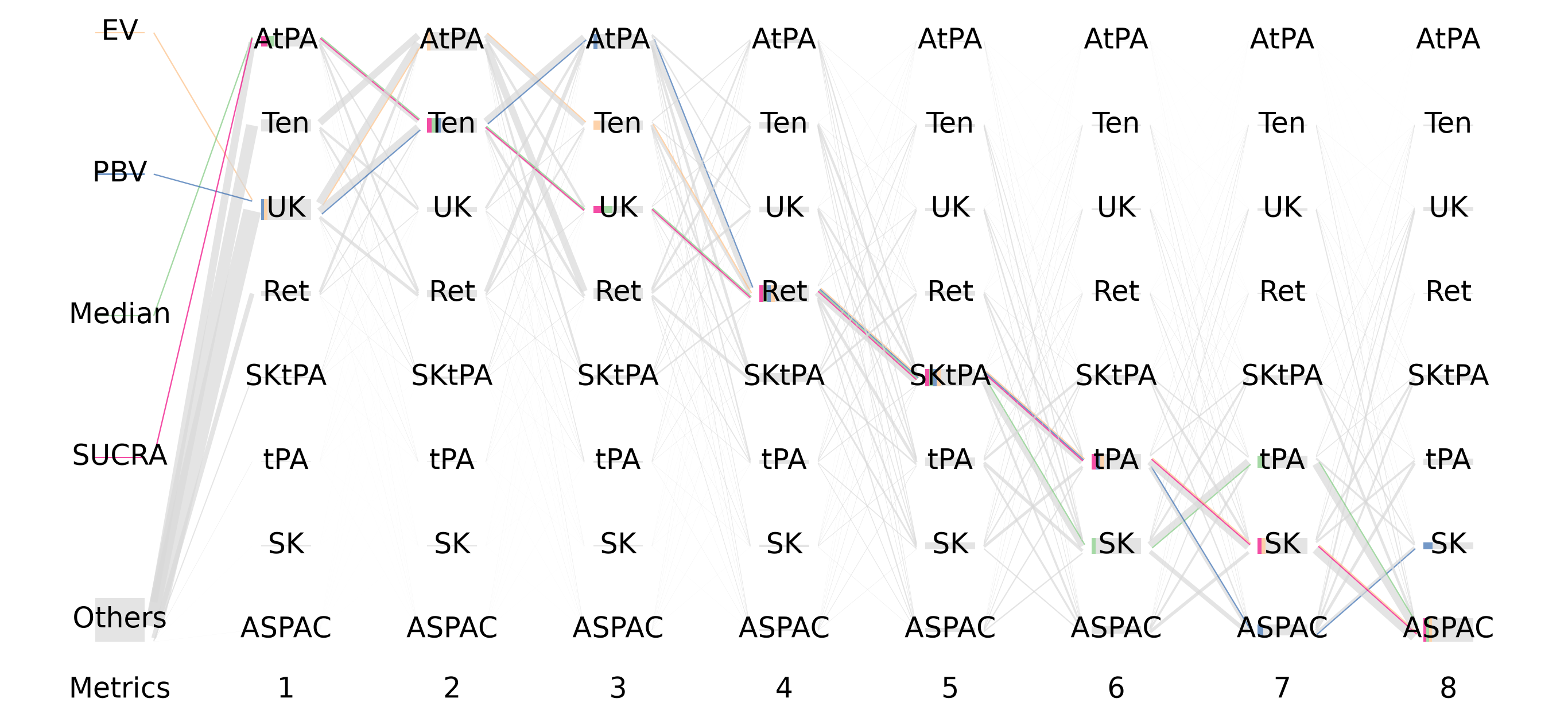}}
    \caption{A hammock plot highlighting ranking metrics for the thrombolytic dataset. Treatments on each rank axis are ordered by SUCRA. To improve visibility, an augmentation is applied to the highlighted hierarchies so that each of them occupies 3\% of the $N$ posterior samples. The thin highlighted paths indicates that hierarchies identified by ranking metrics are not strongly supported by the posterior distribution.}
    \label{fig:Figure10_thrombo_metrics}
\end{figure*}

A consequence of the absence of highly probable treatment hierarchies is that the treatment hierarchy suggested by a ranking metric may be less reliable as a summary of the posterior distribution. Figure~\ref{fig:Figure7_thrombo}D shows that SUCRA, PBV, EV and posterior mean rank identify four different treatment hierarchies. Figure~\ref{fig:Figure10_thrombo_metrics} displays these four hierarchies within the full posterior distribution, and highlights each of them with a different color. 
None of the hierarchies identified by the selected ranking metrics matches any of the three most credible treatment hierarchies in Figure~\ref{fig:Figure9_thrombo_fre}. This does not imply that the metric-based hierarchies are incorrect, because ranking metrics  ``address different ‘treatment hierarchy questions’ based on how the ‘preferable treatment’ is defined'' \citep{chiocchia2023complexity}. However, they receive limited support from the posterior distribution, as indicated by the thin highlighted paths in Figure~\ref{fig:Figure10_thrombo_metrics}, which makes them potentially misleading when interpreted as complete treatment hierarchies. This can occur because ranking metrics typically summarize the marginal ranking distribution of each treatment individually, and combining these summaries into a hierarchy does not necessarily result in a complete treatment hierarchy that is highly supported. Therefore, in this case, the hierarchy selected by a ranking metric should not be interpreted as the definitive treatment ordering, but rather as a summary addressing a particular treatment ranking question with limited posterior support.

Treatment hierarchies based on different ranking metrics may appear consistent even when they are weakly supported. For example, Figure~\ref{fig:Figure7_thrombo}D presents the hierarchies suggested by the four selected ranking metrics for the thrombolytic dataset. Based solely on this visualization, the treatment ordering may appear relatively stable, because the rank assigned to any treatment differs by at most two positions across four metrics. However, such visualizations do not indicate the posterior support for the complete hierarchy. As shown in Figure~\ref{fig:Figure10_thrombo_metrics},  the hierarchies defined by these ranking metrics account for only a small proportion of the posterior ranking distribution. By embedding metric-based hierarchies within the posterior distribution, the highlighted hammock plot allows both the implied treatment ordering and its posterior support to be examined simultaneously.

\subsection{Hammock Plot Reveals Local Structure Within a Globally Uncertain Treatment Hierarchy}

\begin{figure*}[htb]
    \centerline{\includegraphics[width=\linewidth]{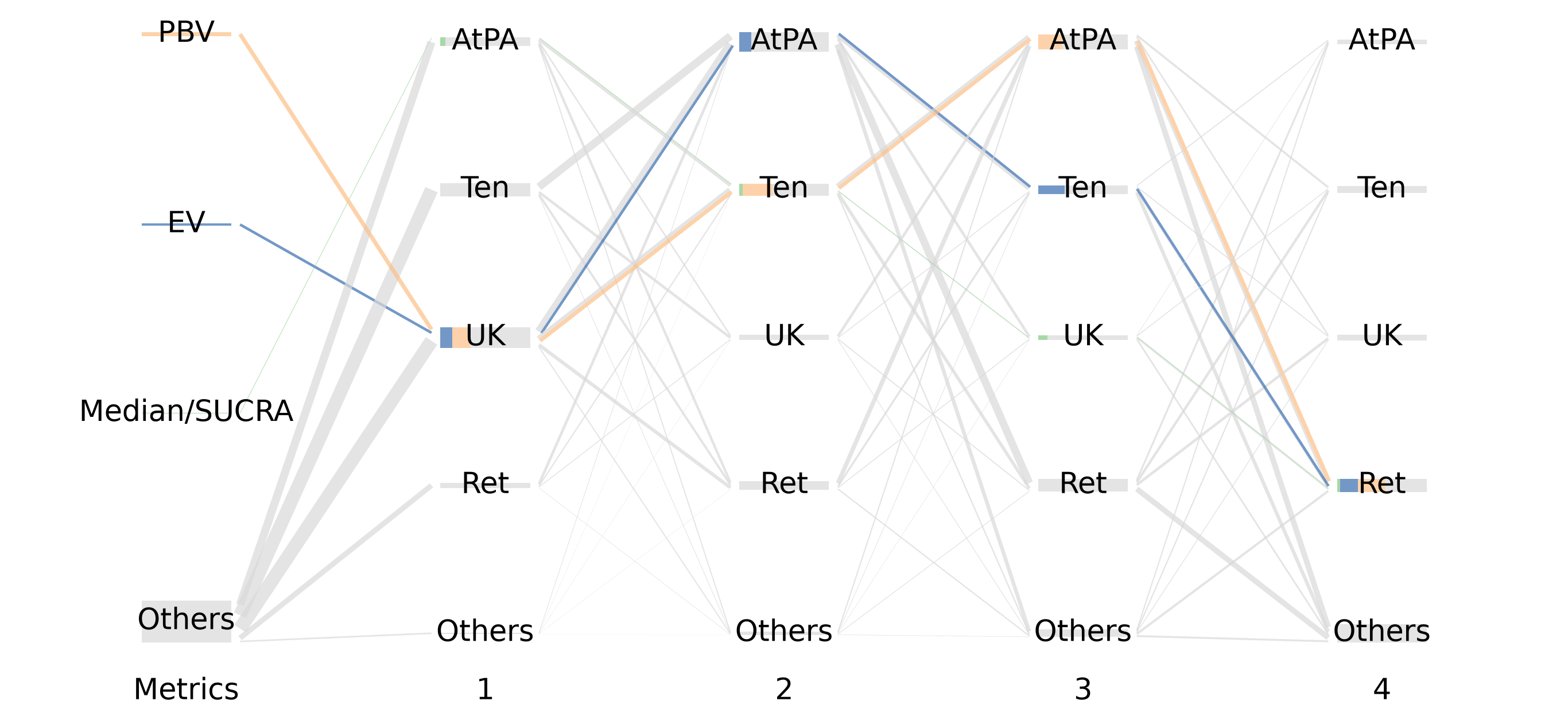}}
    \caption{Top-$k$ hammock plot highlighting ranking metrics for the thrombolytic dataset with $k=4$. The plot focus on the first four ranks and the first four treatments identified by SUCRA (AtPA, Ten, UK and Ret) which correspond to the high-ranked cluster identified before. The PBV and EV paths receive greater posterior support after collapsing information about treatments outside the leading region. Median/SUCRA path remains weakly supported.}
    \label{fig:Figure11_thrombo_top4}
\end{figure*}

The absence of a highly probable complete treatment hierarchy may or may not necessarily imply that all regions of the hierarchy are equally uncertain. The apparent upper-ranked cluster identified in Figure~\ref{fig:Figure8thrombo_simpleHammock} motivates a closer examination of uncertainty within the first four ranks. Figure~\ref{fig:Figure11_thrombo_top4} focuses on the four treatments ranked highest by SUCRA, which are AtPA, Ten, UK and Ret, and they are also the four treatments that we identify as the apparent high-ranked cluster in Section~\ref{sec: treatment_clusters}. The plot focuses on their positions within the original first four ranks, rather than re-ranking these four treatments relative to one another. The remaining treatments are grouped into ``Others", so that posterior samples in which treatments outside this cluster that enter the first four ranks are still represented. 

Although the complete hierarchies identified by the four selected ranking metrics receive extremely weak posterior support, their upper portions do not receive the same level of support. The PBV and EV paths become relatively more prominent when distinctions among the remaining four treatments are collapsed, as indicated by the thicker orange and blue paths in Figure~\ref{fig:Figure11_thrombo_top4}. This suggests that the weak posterior support for their complete hierarchies is partly due to the uncertainty in the ordering of the remaining treatments, since the orderings within the leading region receive considerably greater support. In fact, 8.33\% of the posterior samples support the local ordering identified by PBV and 5.4\% of the samples support that identified by EV. In contrast, the Median/SUCRA path remains relatively thin, with only 1.01\% posterior support. Its weak support in both the complete and local hierarchy indicates the lack of posterior support extends across different portions of the hierarchy. These observations are difficult to obtain from existing ranking visualizations, which generally do not allow posterior support to be examined within a selected region of treatments while preserving information from the full treatment hierarchy.

Within the first three ranks, most connectors are among AtPA, Ten UK and Ret, with only thin connectors to ``Others". This suggests that the first three positions are frequently occupied jointly by three treatments from this four-treatment group, and it occurs in 76.16\% of the posterior samples. These treatments form a group of most effective treatment options, although there remains considerable uncertainty about their relative ordering within the group. The identification of these optimal treatments and the difficulty in distinguishing among them is consistent with previous clinical evidence. The NICE appraisal reported that in terms of 30 day-mortality, Ten and AtPA were considered equivalent \citep{national2002guidance}. For Ret and AtPA, no statistically significant differences in 30-day mortality were observed in the GUSTO-III trial, but the evidence was insufficient to establish equivalence. The evaluation notes that, if Ret was considered equivalent to AtPA, this would indirectly suggest that Ret was as effective as Ten. A similar conclusion was drawn in the threshold analysis by \citet{phillippo2018sensitivity}. UK was not included in these comparisons because it has remained unavailable in the United Kingdom for the use in thrombolytic. The top-$k$ hamock plot highlighting ranking metrics complements these pairwise clinical findings by showing that these treatments tend to appear jointly among the highest ranks within the full eight-treatment hierarchy.

\section{Discussion}\label{s:Discussion}
Existing NMA ranking visualizations primarily focus on marginal ranking distributions or emphasize specific treatment hierarchies identified by ranking metrics. Using hammock plots adapted for use with NMA, we were able to show more detail about the posterior distribution. For the thrombolytic dataset, this has led to additional insights such as broad clusters of higher-ranked and lower-ranked treatments, with substantial ranking uncertainty occurring within each cluster. More generally, hammock plots addressed an important limitation of current approaches in NMA by providing a more comprehensive view of treatment ranking uncertainty and helping to avoid overinterpretation of unstable treatment hierarchies.

In the NMA setting, hammock plot generalized marginal ranking distributions by preserving treatment ranking information while revealing joint posterior behavior. Rank-based and treatment-based hammock plots emphasized different aspects of the posterior distribution, focusing respectively on relationships between adjacent ranks and relative ordering of treatments. We further extended hammock plots to visualize uncertainty in treatment hierarchies through highlighting, which allowed posterior support for specific hierarchies and ranking-metric summaries to be assessed directly. More importantly, highlighting placed selected hierarchies within the broader posterior distribution, allowing them to be interpreted in relation to alternative orderings. Finally, our subsetting approaches enabled focused investigation of local ranking uncertainty while maintaining the context of the full treatment hierarchy. An overall advantage in the design of all hammock plots was direct labeling, i.e. placing labels directly next onto plotting elements rather than using a separate legend, improving readability by allowing treatments or ranks to be followed across the plot without repeatedly referring to a legend.
 
There is a  trade-off between visual complexity and degree of detail that a plot provides. Relative to most other NMA visualizations, hammock plots give a greater amount of detail at the cost of greater visual complexity. However, the additional detail may also lead to additional insights. For example, stacked barcharts provide a simple and intuitive summary of the marginal distribution of ranks for each treatment, but do not show how the rankings of different treatments are related. In contrast, hammock plots retain this marginal information while additionally displaying bivariate relationships and using highlighting to reveal patterns in ranking and hierarchy uncertainty. The choice of visualization should therefore depend on the question of interest. 

Limitations of the proposed methodology include: first, when the number of treatments is very large, the plot may be visually cluttered due to the increasing number of connectors and possible posterior paths. The problem can be mitigated by using top-$k$ hammock plot highlighting ranking metrics or subset hammock plot highlighting ranking metrics. Second, a hammock plot displays only bivariate relationships between neighboring axes rather than the full posterior distribution. However, this goes beyond the marginal distributions examined in the stacked barcharts. Also, highlighting shows the {\em joint} behavior of a limited number of selected paths. Third, the proposed methodology does not incorporate a minimum clinically important difference. This limitation is not specific to hammock plots but is also common to existing NMA ranking visualizations. Fourth, imbalances in the network structure or differences in estimate precision can introduce biases into treatments rankings, and hammock plots will be subject to these biases. This remains an unresolved issue in treatment ranking in NMA rather than a limitation specific to hammock plots.

To facilitate the use of hammock plots in NMA, we have developed a web application implementing all five types of hammock plots, enabling researchers to generate and customize visualizations using their own posterior treatment effect samples without requiring coding expertise and providing nine example NMA datasets for illustration. Beyond NMA, the hammock plots introduced here can be applied to any setting in which samples from a joint ranking distribution are available.

\section*{Acknowledgments}
We gratefully acknowledge funding from the Canadian NSERC Discovery grant program (RGPIN-2019-04404, PI: B{\'e}liveau; RGPIN-2025-00392, PI: Schonlau).

\bibliographystyle{chicago}
\bibliography{arXiv_references} 

\section*{Supplementary Material}
The web application implementing the hammock plots defined in Section~\ref{s:Method} is available at \url{https://hammock-nma.streamlit.app} with source code available at \url{https://github.com/caliangz3/Hammock-NMA}.\vspace*{-8pt}

\clearpage
\appendix
\setcounter{page}{1}
\setcounter{figure}{0}
\setcounter{table}{0}
\renewcommand{\figurename}{Web Figure}
\renewcommand{\tablename}{Web Table}

\section*{Web Appendix}

\subsection*{Web Appendix A. Stacked Barcharts} 
\hypertarget{WebAppendixA}{}

\begin{figure}[h]
    \centering
    \includegraphics[width=1\linewidth]{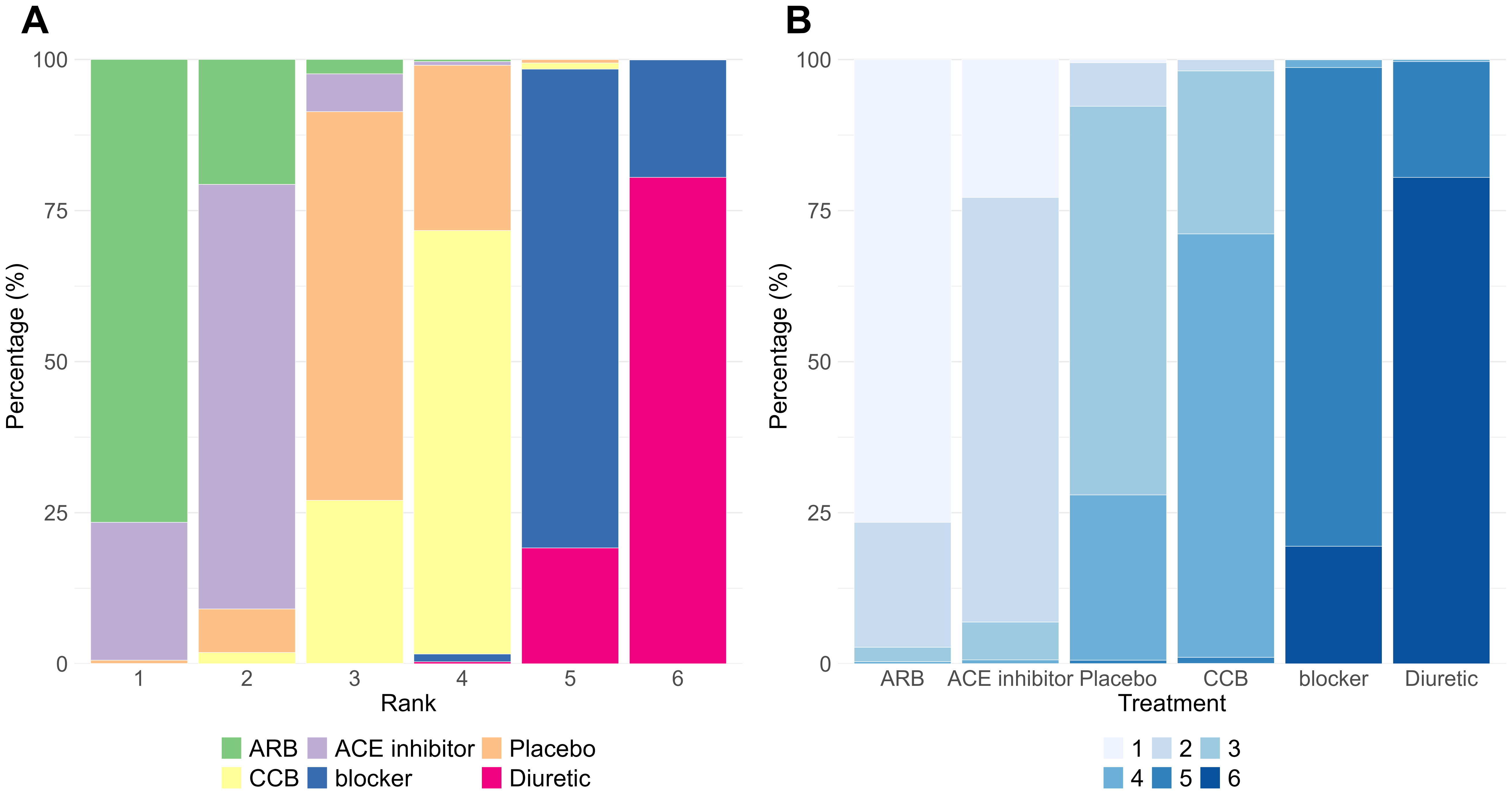}
    \caption{Two equivalent stacked barcharts of the posterior distribution for the diabetes network. (A) Each column presents the marginal posterior probability distribution of treatments at a fixed rank. (B) Each column represents the marginal posterior probability distribution of ranks for a fixed treatment. Treatments are ordered by SUCRA, from top to bottom within each bar of (A) and from left to right across the column labels of (B). In both charts, the height of each segment is proportional to the corresponding posterior probability.}
    \label{fig:Figure1_snapshot}
\end{figure}

In NMA, stacked barcharts are a common way of presenting the marginal posterior probabilities associated with the possible ranks of each treatment. Web Figure~\ref{fig:Figure1_snapshot}
presents two stacked barcharts for the diabetes NMA dataset using these probabilities. In Web Figure~\ref{fig:Figure1_snapshot}A, each bar represents a possible rank, from rank 1 (best) to rank $T$ (worst), while in Web Figure~\ref{fig:Figure1_snapshot}B each bar represents a treatment. Both plots show equivalent information. The height of each stacked segment in a bar is proportional to the posterior probability that treatment $t$ takes rank $r$: $p_{tr}=P\{Treatment(r)=t | data\} = P\{rank(t)=r | data\}.$

Taller segments represent higher posterior probabilities. The absence of a treatment segment on a rank bar in  Web Figure~\ref{fig:Figure1_snapshot}A or a rank segment on a treatment bar in Web Figure~\ref{fig:Figure1_snapshot}B indicates $p_{tr}=0$. For example, Web Figure~\ref{fig:Figure1_snapshot}A shows that treatment ARB has the highest posterior probability of being ranked first, as it occupies the tallest segment in the rank 1 bar. In contrast, treatment Diuretic has zero posterior probability of being ranked first and highest posterior probability of being ranked last, as indicated by the absence of a segment in rank 1 bar and the largest segment in rank 6 bar. Similar conclusions can be drawn from Web Figure~\ref{fig:Figure1_snapshot}B, that is ARB has the highest posterior probability of being ranked first among all possible ranks, while Diuretic has the highest probability of being ranked last among all possible ranks.

For Web Figure~\ref{fig:Figure1_snapshot}A, each bar corresponds to a fixed rank $r$, so the segment heights demonstrate the probability distribution over all treatments and satisfy 
$\sum_{t\in S_T} p_{tr} = 1,$
where $S_T$ is the set of all treatments. In other words, each bar represents the marginal posterior distribution of $Treatment(r)$. Similarly, for Web Figure~\ref{fig:Figure1_snapshot}B, each bar corresponds to a fixed treatment $t$ and represents the marginal posterior distribution of $Rank(t)$, satisfying
$\sum_{r=1}^T p_{tr} = 1.$

Although the two stacked barcharts present the same marginal posterior probabilities from two different perspectives, each summarizes the distribution independently by rank or by treatments. Meanwhile, they do not convey the correspondence between adjacent ranks (or treatments) within the same posterior sample. For example, Web Figure~\ref{fig:Figure1_snapshot}A shows the posterior distribution of treatments at rank 1 and 2 separately. We can estimate $P\{Treatment(1)=\text{ARB}|data\}$ and $P\{Treatment(2)=\text{ACE \ inhibitor}|data\}$ from the heights of the ARB segment in the rank 1 bar and the ACE inhibitor segment in the rank 2 bar, but the two charts do not reveal the joint probability
$P\{Treatment(1)=\text{ARB},~Treatment(2)=\text{ACE \ inhibitor}|data\}$
which describes how two ranks occur together within the same posterior sample.

\newpage
\section*{Web Appendix B. Visualizing a Subset of Treatments} 
\hypertarget{WebAppendixB}{}

When a network consists of a large number of treatments, visualizing its posterior distribution in a hammock plot leads to visual clutter. We propose two approaches in the following sections to reduce visual clutter by subsetting the treatments displayed in a hammock plot. These hammock plots are also naturally useful when specific treatment positions (e.g. top-$k$) or when classes of treatments are of particular interest.

\subsection*{Web Appendix B.1 Subsetting Based on Top-$k$ Treatment Hierarchies}

\begin{figure}[h]
    \centering
    \includegraphics[width=1\textwidth, trim=0 30 0 0, clip]{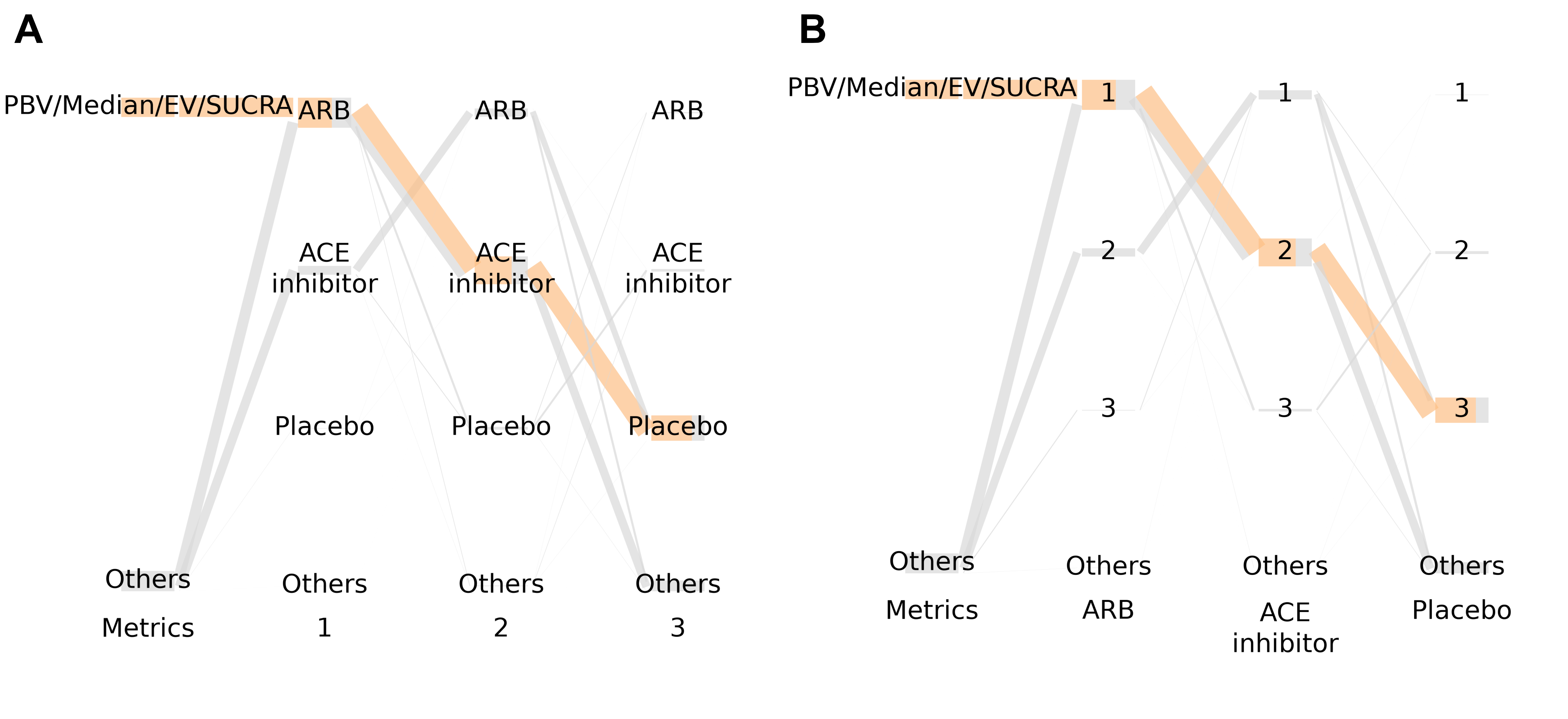}
    \caption{Top-$k$ hammock plot highlighting ranking metics for the treatment path identified by SUCRA ($k=3$) for the diabetes dataset. (A) Each column presents the posterior probability distribution of the selected treatments at a fixed rank no greater than $k$, constructed from the treatment matrix. Treatments with rank greater than $k$ in the hierarchy identified by SUCRA are grouped into ``Others". (B) Each column represents the posterior probability distribution of ranks for the selected treatments, constructed from the rank matrix. Only selected treatment axes are plotted and ranks greater than $k$ are grouped into ``Others". In both charts, highlighted path corresponds to the reduced hierarchies/rankings identified by the corresponding ranking metrics. The width of each highlighted path is proportional to the posterior probability of the corresponding reduced path. Treatments are ordered by SUCRA, from top to bottom within each rank axis of (A) and from left to right across the column labels of (B). Categories on the Metrics axis are ordered by the number of posterior samples in each category, from largest to smallest, with ``Others" always positioned last.}
    \label{fig:Figure5_topK}
\end{figure}

The first subsetting approach focuses on treatments that ranked within the top $k$ positions in a hierarchy identified by a specific ranking metric, where $k < T$. Web Figure~\ref{fig:Figure5_topK} shows two hammock plots after subsetting the diabetes datasets using $k=3$. Only the treatments that occupy the top three positions in the treatment hierarchy identified by SUCRA are illustrated. 

Let $S_k=\{Treatment(1),...,Treatment(k)\}$ denote the set of treatments with a rank no greater than $k$ in the hierarchy identified by a ranking metric. Web Figure~\ref{fig:Figure5_topK}A is constructed using the treatment matrix by replacing all treatments that do not belong to $S_k$ with the category ``Others". Therefore, each parallel rank axis displays the selected treatments individually, and the ``Others" category represents all the remaining treatments. The Metrics axis is constructed similarly as in Figure~\ref{fig:Figure4_metrics}. However, for the hierarchy suggested by each ranking metric, treatments that are not in $S_k$ are also replaced by ``Others" to ensure consistency with the modified treatment matrix. Each modified posterior sample is then assigned the corresponding metric category if the modified sample matches the modified hierarchy identified by that metric. The remaining posterior samples are assigned to the category ``Others".

Let $H_k=[Treatment(1),...,Treatment(k)]$ denote the reduced treatment hierarchy displayed on the first $k$ rank axes. For every rank $r \le k$, $Treatment(r)\in{S_k}\cup\{Others\}$. Since several complete treatment hierarchies can produce the same reduced hierarchy, the width of a highlighted path then represents the sum of the posterior probabilities of all complete treatment hierarchies that can be reduced to the same modified hierarchy. Let $h_{m,k}$ denote the reduced hierarchy generated by metric $m$. The width of its highlighted path is proportional to
$P(H_k = h_{m,k})=\sum_{h:Reduced(h)=h_{m,k}} P(H=h|data),$
where $Reduced(h)$ replaces all treatment outside of $S_k$ with ``Others" in every possible hierarchy $h$. For example, in Web Figure~\ref{fig:Figure5_topK}A, PBV groups all complete hierarchies whose first three treatments are exactly ARB, ACE inhibitor, and Placebo, regardless of how the remaining $T-3$ treatments are ordered. This reduces visual clutter because many complete hierarchies are reduced into fewer paths.

Web Figure~\ref{fig:Figure5_topK}B is constructed similarly to Web Figure~\ref{fig:Figure5_topK}A using the rank matrix. The treatments in $S_k$ are individually displayed as the treatment axes, while all remaining treatments are replaced by ``Others" that are not visualized in the plot. Since $S_k$ contains only the top $k$ treatments suggested by a ranking metrics, samples in the rank matrix with a rank greater that $k$ are also replaced by ``Others". Consequently, each treatment axis consists of the categories $1,...,k$ and $Others$. The same rank replacements are applied to rankings corresponding to the hierarchies identified by ranking metrics. Let $R_k$ denote the reduced ranking, and $r_{m,k}$ denote the reduced ranking identified by metric $m$. The width of the highlighted path is proportional to
$P(R_k=r_{m,k}|data).$
Since many complete rankings also reduce to the same $R_k$, we have
$P(R_k=r_{m,k}|data) = \sum_{r:Reduced'(r)=r_{m,k}} P(R=r|data),$
where $Reduced'(r)$ reduced a complete ranking $r$ by replacing ranks greater than $k$ with ``Others".

\subsection*{Web Appendix B.2 Partial Ordering of Treatments}

The second subsetting approach focuses on a selected subsets of treatments. Treatments outside the selected subsets are completely ignored, including their position in the path. Only the relative ordering of the selected treatments is visualized.

\begin{figure}[h]
    \centering
    \includegraphics[width=1\textwidth, trim=0 0 0 0, clip]{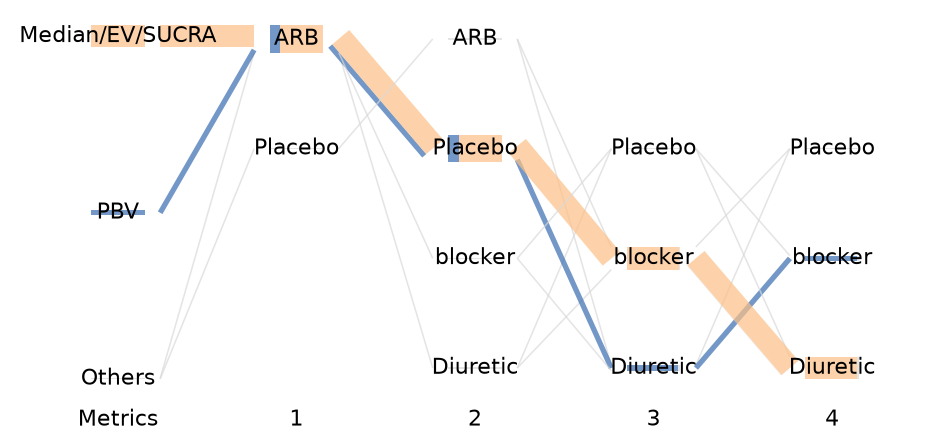}
    \caption{Hammock plot based on a prespecified subset of treatment of the diabetes dataset. The subset of interest is \{ARB, Placebo, blocker, Diuretic\}. Treatment hierarchies are reconstructed by removing all treatments outside the selected subset, while preserving the relative order of the remaining treatments. Therefore, the displayed ranks are relative to the selected subset of treatments rather than the complete treatment hierarchy. The width of the highlighted path of a ranking metrics is proportional to the posterior probability of the corresponding subset hierarchy. Treatments on each rank axis are ordered according to their order in the selected treatment subset. Categories on the Metrics axis are ordered by the number of posterior samples in each category, from largest to smallest, with ``Others" always positioned last.}
    \label{fig:Figure6_subset}
\end{figure}

Web Figure~\ref{fig:Figure6_subset} presents a hammock plot for the diabetes datasets where the subset of interest is \{ARB, Placebo, blocker, Diuretic\}. 
More generally, let $S_{sub}$ denote a prespecified subset of treatments. The plot is generated from a  reconstructed treatment matrix. This treatment matrix is constructed from the posterior samples of the relative treatment effects after removing all treatments not included in $S_{sub}$. For each posterior sample, the remaining treatments are reordered according to their posterior relative treatment effects. This produces treatment hierarchies defined only on $S_{sub}$. Therefore, the displayed ranks are relative to the selected subset rather than the complete treatment hierarchy. For example, if $Rank(t_1)=1$ and $Rank(t_2)=2$ based on the selected subset, this only indicates $Rank(t_1)<Rank(t_2)$ in the completely treatment hierarchy. Similarly, treatments do not belong to $S_{sub}$ are removed from the hierarchies identified by the ranking metrics, leaving only the relative ordering of the selected treatments.

Let $H_{sub}$ denote a treatment hierarchy of the treatments in $S_{sub}$. The width of a highlighted path is proportional to
$P(H_{sub}=h_{sub}|data)$
where $h_{sub}$ is the treatment hierarchy identified by a ranking metric after removing all treatments not in $S_{sub}$. Similar to the first subsetting approach, multiple complete treatment hierarchies may correspond to the same $h_{sub}$. Therefore,
$P(H_{sub}=h_{sub}|data) = \sum_{h:sub(h)=h_{sub}} P(H=h|data),$ 
where $sub(h)$ outputs a hierarchy from $h$ after removing all treatments not in $S_{sub}$, while preserving the relative ordering of the remaining treatments. For example, in Web Figure~\ref{fig:Figure6_subset}, the displayed hierarchy $ARB>Placebo>blocker>Diuretic$ represents every complete treatment hierarchy where these four treatments appear in the same relative order, regardless of the positions of the remaining treatments.

Web Figure~\ref{fig:Figure6_subset} uses a feature of the hammock plot that sets the minimum connector width to 0.2\% of the plot height. This ensures that thin paths with small posterior probabilities are still visible, even though they are not selected by any ranking metrics in this case.

\newpage

\section*{Web Tables}

\begin{table}[h]
\centering
\caption{The first seven posterior samples of relative treatment effects for the diabetes dataset. Smaller relative treatment effects indicate more desirable treatment outcomes.}
\label{tab:relative_effects}
\scalebox{0.97}{
\begin{tabular}{rrrrrr}
\toprule
Diuretic & ACE inhibitor  &  ARB  & blocker & CCB & Placebo   \\
\midrule
0   &   -0.36103 & -0.47534 & -0.02478 & -0.28124 & -0.23984 \\
0   &   -0.29515 & -0.56692 & 0.00384 & -0.21582  & -0.21214 \\
0   &   -0.32656 & -0.71906 & -0.01927 & -0.26625 & -0.22171 \\
0   &   -0.37795 & -0.50719 & -0.11166 & -0.28962 & -0.17332 \\
0   &   -0.51065 & -0.67437 & -0.15898 & -0.34252 & -0.35347 \\
0   &   -0.44904 & -0.59989 & -0.19297 & -0.35244 & -0.26743 \\
0   &   -0.49931 & -0.54690 & -0.14725 & -0.29033 & -0.24204 \\
\bottomrule
\end{tabular}}
\end{table}

\begin{table}[h]
\centering
\caption{Rank matrix corresponding to the first seven posterior samples of relative treatment effects for the diabetes dataset.}
\label{tab:rank_matrix}
\scalebox{0.97}{
\begin{tabular}{rrrrrr}
\toprule
Diuretic & ACE inhibitor  &  ARB  & blocker & CCB & Placebo   \\
\midrule
6      &       2  & 1    &   5 &  3    &   4 \\
5      &       2  & 1    &   6 &  3    &   4 \\
6      &       2  & 1    &   5 &  3    &   4 \\
6      &       2  & 1    &   5 &  3    &   4 \\
6      &       2  & 1    &   5 &  4    &   3 \\
6      &       2  & 1    &   5 &  3    &   4 \\
6      &       2  & 1    &   5 &  3    &   4 \\
\bottomrule
\end{tabular}}
\end{table}

\begin{table}[h]
\centering
\caption{Treatment matrix corresponding to the first seven posterior samples of relative treatment effects for the diabetes dataset.}
\label{tab:treatment_matrix}
\scalebox{0.97}{
\begin{tabular}{llllll}
\toprule
1 & 2  &  3  & 4 & 5 & 6   \\
\midrule
ARB & ACE inhibitor & CCB  &    Placebo & blocker &  Diuretic \\
ARB & ACE inhibitor & CCB  &    Placebo & Diuretic & blocker  \\
ARB & ACE inhibitor & CCB  &    Placebo & blocker &  Diuretic \\
ARB & ACE inhibitor & CCB  &    Placebo & blocker &  Diuretic \\
ARB & ACE inhibitor & Placebo & CCB   &   blocker &  Diuretic \\
ARB & ACE inhibitor & CCB  &    Placebo & blocker &  Diuretic \\
ARB & ACE inhibitor & CCB  &    Placebo & blocker &  Diuretic \\
\bottomrule
\end{tabular}}
\end{table}

\newpage

\begin{table}[ht]
\raggedright
\caption{An overview of hammock plot visualization methods for NMA.}
\label{tab:hammock_plots}
\begin{tabularx}{\linewidth}{
  @{}
  >{\raggedright\arraybackslash}p{4cm}
  >{\raggedright\arraybackslash}X
  c
  l
  @{}
}
\hline
Name & Description & Highlighting & Example \\
\hline
Standard hammock plot & 
Extends the stacked barcharts by visualizing pairwise distribution of adjacent treatments or ranks while preserving marginal distribution of all rank or all treatments & 
no&
Figure~\ref{fig:Figure2_simpleHammock}\\

Hammock plot highlighting frequency&
Extends the standard hammock plot by highlighting the most frequent treatment hierarchies or rankings to reveal their corresponding posterior probabilities and the uncertainty in the joint posterior distribution &
yes&
Figure~\ref{fig:Figure3_Frequency}\\

Hammock plot highlighting ranking metrics &
Extends the standard hammock plot by highlighting the treatment hierarchies identified by different ranking metrics to assess how well these hierarchies are supported by the posterior distribution and the associated uncertainty &
yes &
Figure~\ref{fig:Figure4_metrics}\\

Top-$k$ hammock plot highlighting ranking metrics &
Focuses on the top $k$ rank positions and the corresponding treatments identified by a ranking metric. All remaining treatments are grouped as ``Others" treatment category, while ranks greater than $k$ are grouped as ``Others" rank category. These groupings help to reduce the number of distinct sample paths. Hierarchies (or rankings) identified by ranking metrics are highlighted &
yes &
Web Figure~\ref{fig:Figure5_topK} \\

Subset hammock plot highlighting ranking metrics &
Focuses on a selected subset of treatments and their relative ordering by excluding other treatments, and helps to reduce the number of distinct samples paths. Hierarchies identified by ranking metrics are highlighted. &
yes &
Web Figure~\ref{fig:Figure6_subset} \\
\hline

\end{tabularx}
\end{table}

\newpage

\section*{Web Figures}
\begin{figure}[h]
    \centering
    \includegraphics[width=\linewidth, trim=30 50 20 40, clip]{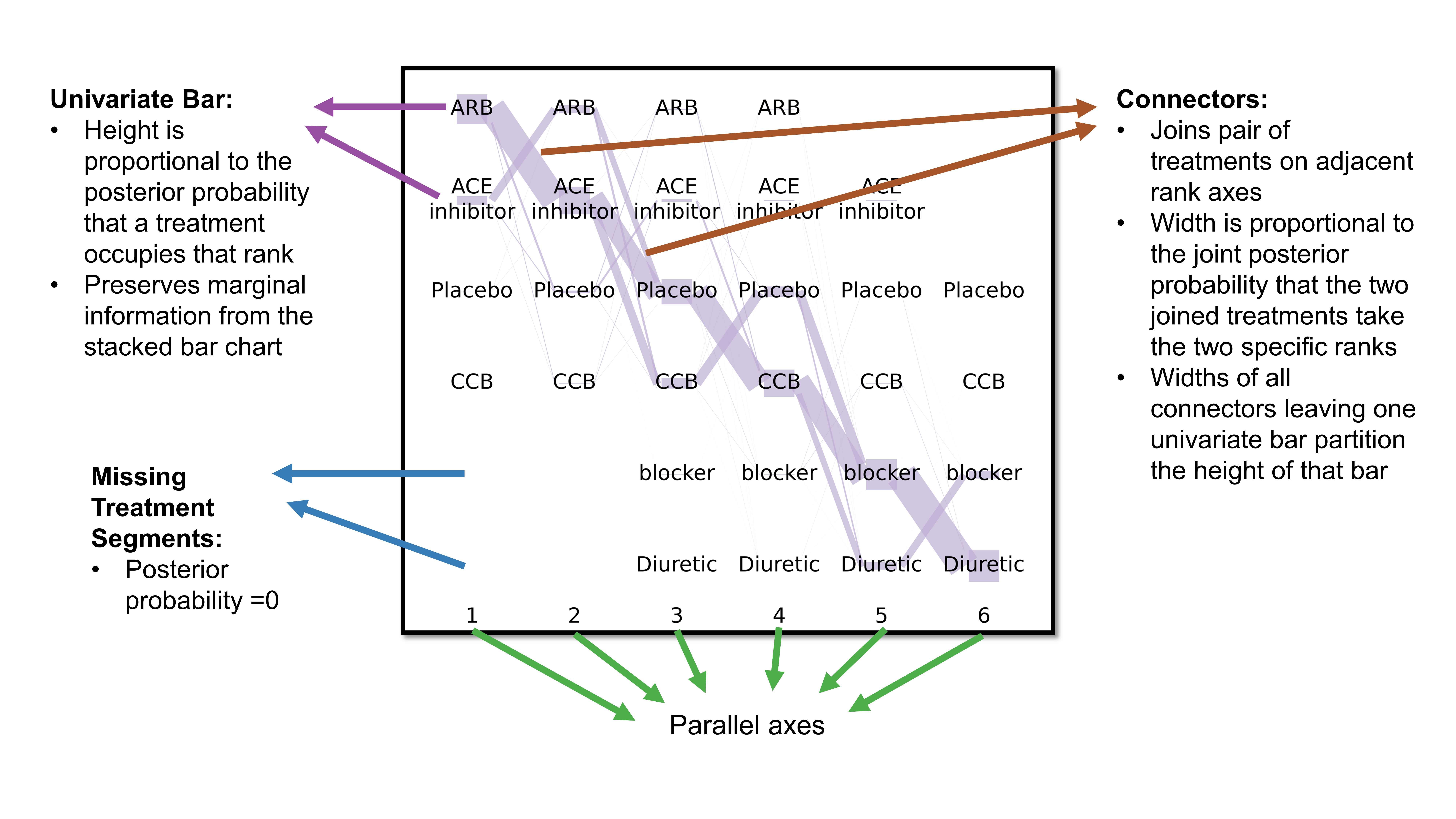}
    \caption{Annotated version of Figure~\ref{fig:Figure2_simpleHammock}A illustrating the main graphical elements of a hammock plot. The annotations identify each element and explain what they represent.}
    \label{fig:Figure12_Hammock_annotation}
\end{figure}

\end{document}